\documentclass[sigconf,twocolumn,nonacm]{acmart}
\AtBeginDocument{%
  \providecommand\BibTeX{{%
    Bib\TeX}}}
    
\setcopyright{acmlicensed}
\copyrightyear{2026}
\acmYear{2026}
\acmDOI{XXXXXXX.XXXXXXX}
\acmConference[SBC '26]{Science of Blockchain '26}{July 27--29,
  2026}{CA, USA}
\acmISBN{978-1-4503-XXXX-X/2018/06}

\usepackage{tikz}
\usepackage{amsmath}
\AtBeginDocument{%
  \providecommand\BibTeX{{%
    Bib\TeX}}}

\usepackage[most]{tcolorbox} % TODO check whether allowed by CCS, breaks with the default empty parameters
\usepackage{algorithmic}
\usepackage[binary-units=true]{siunitx}
\usepackage{graphicx}
\usepackage{glossaries}
\usepackage{textcomp}
\usepackage[dvipsnames]{xcolor} % FIXME ideally would use this options if allowed by CCS instead of the default version below
\usepackage{hyperref}
\usepackage{cleveref}
\usepackage{booktabs}
\usepackage[shortlabels]{enumitem}
\usepackage{makecell}
\usepackage{subcaption}
\usepackage{multirow}
\usepackage{tikz}
\usepackage{amsfonts}
\usepackage[english]{babel}
\usepackage[T1]{fontenc}
\usepackage{xspace}
\usepackage{algorithm2e}

\usetikzlibrary{positioning,shapes,arrows,calc}
\usepackage{pifont} % for \ding checkmarks

\def\BibTeX{{\rm B\kern-.05em{\sc i\kern-.025em b}\kern-.08em
    T\kern-.1667em\lower.7ex\hbox{E}\kern-.125emX}}

\DeclareSIUnit[per-mode=symbol]\bps{\bit\per\second}
\DeclareSIUnit[per-mode=symbol]\kbps{\kilo\bps}
\DeclareSIUnit[per-mode=symbol]\Mbps{\mega\bps}
\DeclareSIUnit[per-mode=symbol]\Gbps{\giga\bps}
\DeclareSIUnit[per-mode=symbol]\nanosec{\nano\second}
\DeclareSIUnit[per-mode=symbol]\packet{packet}
\DeclareSIUnit[per-mode=symbol]\packetps{\packet\per\second}
\DeclareSIUnit\microsec{\SIUnitSymbolMicro s}
\DeclareSIUnit\byte{B}
\DeclareSIUnit\bit{bit}
\DeclareSIUnit\terabyte{TB}

\makeglossaries

\newtheorem{definition}{Definition}

\usetikzlibrary{arrows}

\SetKwProg{fn}{function}{}{end}
\SetKw{require}{require}
\newcommand{\mc}{\mathcal}
\newcommand{\msf}{\mathsf}

\newcommand{\ie}{i.e., }

\newcommand{\secp}{\ensuremath{\lambda}}
\newcommand{\secparam}{\secp}

\newcommand{\adv}{\ensuremath{\mathcal{A}}}

\newcommand{\msg}[2]{%
\ifthenelse{\isempty{#1}}{\ensuremath{(#2)}}{%
\ifthenelse{\isempty{#2}}{\ensuremath{(\mathsf{#1})}}{%
\ensuremath{(\mathsf{#1}, #2)}}}}

\newcommand{\tee}{\msf{TEE}}
\newcommand{\satellite}{\mc{S}}
\newcommand{\satelliteSet}{\mathbb{S}}
\newcommand{\gs}{\mc{GS}}
\newcommand{\gsSet}{\mathbb{GS}}
\newcommand{\totalGS}{n}
\newcommand{\adversarialGSPercentage}{t}
\newcommand{\adversarialGS}{t_\text{GS}}
\newcommand{\adversarialChannels}{t_\text{ch}}
\newcommand{\channelCorruptionWindow}{W}

\newcommand{\tenant}{\mc{T}}

\newcommand{\keygen}{\msf{KeyGen}}
\newcommand{\sign}{\msf{Sign}}
\newcommand{\verify}{\msf{Verify}}

\newcommand{\privkey}{\msf{sk}}
\newcommand{\pubkey}{\msf{vk}}
\newcommand{\signature}{\sigma}

\newcommand{\teeProof}{\Pi_\msf{TEE}}
\newcommand{\timestamp}{\msf{ts}}

\newcommand{\certificate}{C}

\newcommand{\nonce}{\phi}

\newcommand{\availabilityParam}{u}

\begin{document}
\newacronym{tee}{TEE}{Trusted Execution Environment}
\newacronym{tpm}{TPM}{Trusted Platform Module}
\newacronym{hsm}{HSM}{Hardware Security Module}
\newacronym{tcb}{TCB}{Trusted Computing Base}
\newacronym{hw}{HW}{Hardware}
\newacronym{sw}{SW}{Software}
\newacronym{l1}{L1}{Layer-1}
\newacronym{l2}{L2}{Layer-2}
\newacronym{vm}{VM}{Virtual Machine}
\newacronym{da}{DA}{Data Availability}
\newacronym{defi}{DeFi}{Decentralized Finance}
\newacronym{bft}{BFT}{Byzantine Fault Tolerant}
\newacronym{tps}{TPS}{transaction per second}
\newacronym{bridge-contract}{bridge-contract}{bridge-contract}
\newacronym{fpf}{fast pre-finality}{fast pre-finality}
\newacronym{dcea}{DCEA}{Data Center Execution Assurance}
\newacronym{seap}{SEAP}{Satellite Execution Assurance Protocol}
\newacronym{sea}{SEA}{Satellite Execution Assurance}
\newacronym{kms}{KMS}{Key Management Service}
\newacronym{mee}{MEE}{Memory Encryption Engine}
\newacronym{tzasc}{TZASC}{TrustZone Address Space Controller}
\newacronym{tzpc}{TZPC}{TrustZone Protection Controller}
\newacronym{rme}{RME}{Realm Management Extension}
\newacronym{rmm}{RMM}{Realm Management Monitor}
\newacronym{gpc}{GPC}{Granule Protection Check}
\newacronym{el}{EL}{Exception Level}
\newacronym{srk}{SRK}{Storage Root Key}
\newacronym{leo}{LEO}{Low Earth Orbit}
\newacronym{geo}{GEO}{Geostationary Earth Orbit}
\newacronym{ta}{TA}{Trusted Application}
\newacronym{ra}{RA}{Remote Attestation}
\newacronym{gs}{GS}{Ground Station}
\newacronym{rot}{RoT}{Root of Trust}
\newacronym{hab}{HAB}{High Assurance Boot}
\newacronym{rpmb}{RPMB}{Replay Protected Memory Block}
\newacronym{mitm}{MitM}{Man-in-the-Middle}
\newacronym{cpak}{CPAK}{Platform Attestation Key}
\newacronym{crk}{CRK}{Creator Root Key}
\newacronym{rpk}{RPK}{Root Provisioning Key}
\newacronym{crt}{CRT}{Create Root Key}
\newacronym{otp}{OTP}{One Time Programmable}
\newacronym{trng}{TRNG}{True Random Number Generator}
\newacronym{eat}{EAT}{Entity Attestation Token}
\newacronym{pqc}{PQC}{Post Quantum Cryptography}
\newacronym{client}{CLNT}{client}

% \glsadd{bridge-contract}
\newacronym{nic}{NIC}{Network Interface Card}
\newacronym{iiot}{IIoT}{Industrial Internet of Things}
\newacronym{iot}{IoT}{Internet of Things}
\newacronym{cots}{COTS}{Commercial off-the-Shelf}
\newacronym{rtt}{RTT}{Round Trip Time}
\newacronym{e2e}{E2E}{End-to-End}
\newacronym{p2p}{P2P}{Peer-to-Peer}
\newacronym{gptp}{gPTP}{generic Precision Time Protocol}
\newacronym{ntp}{NTP}{Network Time Protocol}
\newacronym{phc}{PHC}{PTP Hardware Clock}
\newacronym{gm}{GM}{Grandmaster Clock}
\newacronym{tc}{tc}{traffic control}
\newacronym[plural=TCLs,firstplural=traffic classes (TCLs)]{tcl}{TCL}{Traffic Class}
\newacronym{qos}{QoS}{Quality of Service}
\newacronym{ecdf}{ECDF}{Empirical Cumulative Distribution Function}
\newacronym{be}{BE}{Best Effort}
\newacronym{kpi}{KPI}{Key Performance Indicator}
\newacronym{skb}{SKB}{Socket Buffer}
\newacronym{sut}{SUT}{System Under Test}
\newacronym{phy}{PHY}{Physical Layer}
\newacronym{udp}{UDP}{User Datagram Protocol}
\newacronym{bmca}{BMCA}{Best Master Clock Algorithm}
\newacronym{tcp}{TCP}{Transmission Control Protocol}
\newacronym{os}{OS}{Operating System}
\newacronym{irq}{IRQ}{Interrupt Request}
\newacronym{cpu}{CPU}{Central Processing Unit}
\newacronym{smp}{SMP}{Symmetrical Multiprocessing}
\newacronym{smt}{SMT}{Simultaneous Multi-Threading}
\newacronym{rt}{RT}{Real-Time}
\newacronym{kc}{KC}{Key Contribution}
\newacronym{pcap}{PCAP}{Packet Capture}
\newacronym{utc}{UTC}{Coordinated Universal Time}
\newacronym{tai}{TAI}{International Atomic Time}
\newacronym{txtime}{TxTime}{transmission time}
\newacronym{macsec}{MACsec}{Media Access Control Security}
\newacronym{hpc}{HPC}{High Performance Computer}
\newacronym{lpc}{LPC}{Low Performance Computer}
\newacronym{engine}{EnGINE}{Environment for Generic In-vehicular Networking Experiments}
\newacronym{sc}{SC}{Secure Channel}
\newacronym{se}{SE}{Secure Element}
\newacronym{pmp}{PMP}{Physical Memory Protection}
\newacronym{fpga}{FPGA}{Field Programmable Gate Array}
\newacronym{fqcodel}{FQ\_CoDel}{Fair Queuing with Controlled Delay}
\newacronym{mac}{MAC}{Media Access Control}
\newacronym{ip}{IP}{Internet Protocol}
\newacronym{ws}{WS}{window size}
\newacronym{ram}{RAM}{Random-Access Memory}
\newacronym{is}{IS}{Interframe Spacing}
\newacronym{god}{GOD}{Guaranteed Output Delivery}
\newacronym{pp}{PP}{Payment Processor}
\newacronym{dlp}{DLP}{Discrete Logarithm Problem}
\newacronym{saas}{SaaS}{Software-as-a-Service}
\newacronym{dsa}{DSA}{Digital Signature Algorithm}
\newacronym{ecdsa}{ECDSA}{Elliptic Curve Digital Signature Algorithm}
\newacronym{eddsa}{EdDSA}{Edwards-curve DSA}
\newacronym{dkg}{DKG}{Distributed Key Generation}
\newacronym{zkp}{ZKP}{Zero-Knowledge Proof}
\newacronym{PoC}{PoC}{proof-of-concept}
\newacronym{mpc}{MPC}{Multiparty Computation}
\newacronym{ot}{OT}{Oblivious Transfer}
\newacronym{vCPUs}{vCPUs}{virtual CPUs}
\newacronym{txt}{TXT}{Trusted Execution Technology}
\newacronym{tdx}{TDX}{Trust Domain Extensions}
\newacronym{sev}{SEV}{Secure Encrypted Virtualization}
\newacronym{snp}{SNP}{Secure Nested Paging}
\newacronym{sgx}{SGX}{Software Guard Extensions}
\newacronym{tz}{TZ}{TrustZone}
\newacronym{cca}{CCA}{Confidential Computing Architecture}
\newacronym{qemu}{QEMU}{Quick Emulator}
\newacronym{kvm}{KVM}{Kernel-based Virtual Machine}
\newacronym{tsn}{TSN}{Time Sensitive Networking}
\newacronym{methoda}{METHODA}{Multilayer Environment and Toolchain for Holistic NetwOrk Design and Analysis}
\newacronym{pos}{pos}{plain orchestrating service}
\newacronym{dma}{DMA}{Direct Memory Access}
\newacronym{td}{TD}{Trust Domain}
\newacronym{seam}{SEAM}{Secure Arbitration Mode}
\newacronym{maccode}{MAC}{Message Authentication Code}
\newacronym{gcp}{GCP}{Google Cloud Platform}
\newacronym{sme}{SME}{AMD Secure Memory Encryption}
\newacronym{asp}{ASP}{AMD Secure Processor}
\newacronym{psp}{PSP}{AMD Platform Security Processor}
\newacronym{es}{ES}{Encrypted State}
\newacronym{sota}{SotA}{State of the Art}
\newacronym{poc}{PoC}{Proof of Concept}
\newacronym{vlek}{VLEK}{Verified Launch Enclave Key}
\newacronym{vcek}{VCEK}{Verified Chip Endorsement Key}
\newacronym{vmrk}{VMRK}{VM Root Key}
\newacronym{kds}{KDS}{Key Distribution Server}
\newacronym{ark}{ARK}{AMD Root Key}
\newacronym{ask}{ASK}{AMD SEV Key}
\newacronym{qgs}{QGS}{quote generation service}
\newacronym{pccs}{PCCS}{Provisioning Certification Caching Service}
\newacronym{pcs}{PCS}{Intel Provisioning Certification Service}
\newacronym{mpa}{MPA}{Multi-package Registration Agent}
\newacronym{mktme}{MKTME}{Multi-key Total Memory Encryption}
\newacronym{vmm}{VMM}{VM Manager}
\newacronym{tls}{TLS}{Transport Layer Security}
\newacronym{pek}{PEK}{Platform Endorsement Key}
\newacronym{csr}{CSR}{Certificate Signing Request}
\newacronym{pckcert}{PCKC}{Provisioning Certification Key Certificate}
\newacronym{pce}{PCE}{Provisioning Certificate Enclave}
\newacronym{pck}{PCK}{Provisioning Certification Key}
\newacronym{qsk}{QSK}{Quote Signing Key}
\newacronym{tdqe}{TDQE}{TD Quoting Enclave}
\newacronym{qe}{QE}{TD Quoting Enclave}
\newacronym{ak}{AK}{Attestation Key}
% \newacronym{svn}{SVN}{Security Version Number}
\newacronym{r3aal}{R3AAL}{Ring3 Attestation Abstraction Library}
\newacronym{tdqd}{TDQD}{TD Quote Driver}
\newacronym{qgl}{QGL}{Quote Generation Library}
\newacronym{qvl}{QVL}{Quote Verification Library}
\newacronym{qve}{QVE}{Quote Verification Enclave}
\newacronym{crl}{CRL}{Certificate Revocation List}
\newacronym{spd}{SPD}{Seriel Presence Detect}
\newacronym{rmp}{RMP}{Reverse Map Table}
\newacronym{xts}{XTS}{XEX-based Tweaked CodeBook Mode with Ciphertext Stealing}
\newacronym{xex}{XEX}{XOR-Encrypt-XOR}
\newacronym{aes}{AES}{Advanced Encryption Standard}
\newacronym{zk}{ZK}{Zero-knowledge}
\newacronym{ecc}{ECC}{Elliptic Curve Cryptography}
\newacronym{rsa}{RSA}{Rivest–Shamir–Adleman}
\newacronym{nist}{NIST}{National Institute of Standards and Technology}
\newacronym{bls}{BLS}{Boneh-Lynn-Shacham}
\newacronym{smi}{SMI}{System Management Interrupt}
\newacronym{smm}{SMM}{System Management Mode}
\newacronym{vmpl}{VMPL}{Virtual Machine Protection Level}
\newacronym{mmio}{MMIO}{Memory-mapped I/O}
\newacronym{svsm}{SVSM}{Secure VM Service Module}
\newacronym{vtpm}{vTPM}{Virtual TPM}
\newacronym{vmx}{VMX}{Virtual Machines Extension}
\newacronym{cvm}{CVM}{Confidential VM}
\newacronym{tcg}{TCG}{Trusted Computing Group}
\newacronym{rtmr}{RTMR}{Runtime Extendable Measurement Register}
\newacronym{mr}{MR}{Measurement Register}
\newacronym{mrtd}{MRTD}{Measurement of Trust Domain}
\newacronym{ppid}{PPID}{Protected Platform Identifier}
\newacronym{uuid}{UUID}{Universally Unique Identifier}
\newacronym{pki}{PKI}{Public Key Infrastructure}
\newacronym{as}{AS}{Autonomous System}
\newacronym{vtl}{VTL}{Virtual Trust Level}
\newacronym{dcap}{DCAP}{Data Center Attestation Primitives}
\newacronym{mbr}{MBR}{Master Boot Record}
\newacronym{pcr}{PCR}{Platform Configuration Register}
\newacronym{vt}{VT}{Virtualization Technology}
\newacronym{crtm}{CRTM}{Core Root of Trust for Measurement}
\newacronym{ek}{EK}{Endorsement Key}
\newacronym{ekc}{EKC}{EK Certificate}
\newacronym{drtm}{DRTM}{Dynamic Root of Trust Measurement}
\newacronym{mle}{MLE}{Measured Launch Environment}
\newacronym{acm}{ACM}{Authenticated Code Module}
\newacronym{aws}{AWS}{Amazon Web Services}
\newacronym{svm}{SVM}{Secure Virtual Machine}
\newacronym{tdvf}{TDVF}{Trust Domain firmware}
\newacronym{ct}{CT}{Certificate Transparency}
\newacronym{poe}{POE}{Platform Ownership Endorsement}
\newacronym{guc}{GUC}{Global Universal Composition}
\newacronym{iti}{ITI}{Turing Machine Instance}
\newacronym{ca}{CA}{Certificate Authority}
\newacronym{uc}{UC}{Universal Composition}
\newacronym{puf}{PUF}{Physical Unclonable Function}
\newacronym{svn}{SVN}{Space Vehicle Number}
\newacronym{pta}{PTA}{Pseudo TA}
\newacronym{poet}{Proof of ET}{Proof of Execution Triangulation}
\newacronym{tid}{TID}{Total Ionizing Dose}
\newacronym{see}{SEE}{Single Event Effect}
\newacronym{seu}{SEU}{Single Event Upset}
\newacronym{sel}{SEL}{Single Event Latch-up}

\newcommand{\encircled}[2][0.8mm]{%
    \raisebox{.5pt}{%
        \textcircled{%
            \raisebox{0.35pt}{%
                \kern #1
                \scalebox{0.70}{#2}
            }%
        }%
    }%
}
\makeatletter
\newcommand*{\ensquared}[1]{\relax\ifmmode\mathpalette\@ensquared@math{#1}\else\@ensquared{#1}\fi}
\newcommand*{\@ensquared@math}[2]{\@ensquared{$\m@th#1#2$}}
\newcommand*{\@ensquared}[1]{%
\tikz[baseline,anchor=base]{\node[draw,outer sep=0pt,inner sep=0.6mm,minimum width=4mm] {#1};}} 
\makeatother

\definecolor{ourgreen}{rgb}{0.00,0.49,0.19}
\definecolor{ourred}{rgb}{0.77,0.03,0.11}
\definecolor{ourorange}{rgb}{0.89,0.45,0.13}
\definecolor{ourgrey}{rgb}{0.60,0.60,0.60}
\def\yes{\textcolor{ourgreen}{\large\checkmark}}
\def\maybe{\textcolor{ourorange}{\Large$\circ$}} % $\mathbit
\def\no{\textcolor{ourred}{\Large\texttimes}}
\def\unknown{\textcolor{ourgrey}{\encircled[1mm]{?}}}
\title{Space Fabric: A Satellite-Enhanced Trusted Execution Architecture}

\author{Filip Rezabek}
\affiliation{%
  \institution{SpaceComputer}
  \city{}
  \country{}
  }
\email{filip@spacecomputer.io}

\author{Dahlia Malkhi}
\affiliation{%
  \institution{UCSB \& SpaceComputer}
  % \city{Santa Barbara}
  \country{}
  }
\author{Amir Yahalom}
\affiliation{%
  \institution{SpaceComputer}
  % \city{Munich}
  \country{}
  }

\renewcommand{\shortauthors}{Rezabek et al.}

% \received{14 January 2026}
% \received[revised]{12 March 2026}

\begin{abstract}
The emergence of decentralized satellite networks and orbital computing platforms creates a pressing need for trust architectures that can operate without physical access to the hardware, without reliance on pre-provisioned vendor secrets, and without dependence on a single manufacturer's attestation service. Terrestrial Trusted Execution Environments (TEEs) are insufficient for this setting: hardware-based designs are susceptible to physical attacks. More fundamentally, most of the current platforms root their attestation chains in secrets provisioned during manufacturing, creating a pre-launch trust window and single-vendor dependency that cannot be independently audited.

We present Space Fabric, an architecture that provides the missing trust foundation for orbital computing by relocating the trusted computing stack to satellite infrastructure, exploiting a satellite's post-launch physical inaccessibility as a tamper barrier unattainable by terrestrial deployments.
Our Satellite Execution Assurance Protocol binds workload execution to a specific satellite via a Byzantine-tolerant endorsement quorum of distributed ground stations, certifying not only \emph{what} program executes inside the TEE but also \emph{where}.
A further contribution is fully on-orbit key generation: cryptographic secrets are generated within co-located secure elements after launch, with no persistent signing secrets accessible on Earth at any point.
The residual trust assumption reduces to the operator correctly registering device identifiers, a standard operational requirement common to TEE deployments.
To further reduce single-vendor dependence, Space Fabric distributes its hardware trust anchor across two independent secure elements, an NXP SE050 and a TROPIC01, both of which must co-sign attestation evidence. We implement Space Fabric on a USB Armory Mk II with ARM TrustZone, verify attestation end-to-end using Veraison, and provide a security analysis establishing both satisfaction arguments and impossibility bounds under a strong adaptive adversary.

\end{abstract}
\keywords{TEE, Space, Security, TPM, SEAP, HSM}
% \received[accepted]{5 June 2026}

\maketitle

\section{Introduction}\label{sec:introduction}
The space domain is undergoing a fundamental transformation. 
The proliferation of low-cost small satellites and commercial launch services has made orbital infrastructure increasingly accessible, opening new possibilities not only for traditional space missions but also for general-purpose computation~\cite{feilden2024starcloud,aguerayarcas2025suncatcher,ascend2024}.
Recent proposals for Decentralized Satellite Networks envision constellations where multiple operators share orbital resources, pooling communication, storage, and processing capabilities across organizational boundaries~\cite{seoyual2024hotnets}. 
The trend towards AI further motivates the development of space data centers that can harness solar energy efficiently and benefit from passive thermal dissipation~\cite{feilden2024starcloud,aguerayarcas2025suncatcher,ascend2024}. 
Early use cases — secure multi-tenant edge computing, sovereign data processing, on-orbit \gls{kms}, and verifiable scientific data provenance — are already within reach of current small satellite platforms.

Yet this shift introduces a critical unsolved problem: \emph{how does a tenant or verifier establish trust in computation executing on orbital infrastructure they do not own, cannot physically inspect, and cannot independently monitor?}
In terrestrial cloud computing, this question is addressed by \glspl{tee}, which isolate sensitive code and data from the rest of the system, including privileged software such as the operating system or hypervisor, promising strong confidentiality and integrity guarantees even on untrusted platforms~\cite{intel2023tdx,amd2020sevsnp}.
We differentiate between two main families of \glspl{tee}: process-based and \gls{vm}-based solutions.
The main difference lies in the size of the \gls{tcb}: process-based designs run only specific sensitive operations within the enclave, whereas \gls{vm}-based solutions place the entire guest OS into the enclave.
In practice, this translates into a trade-off between security and usability, as \gls{vm}-based solutions do not require large code modifications and fit current cloud deployments.
\gls{tee} features have driven widespread adoption: Intel \gls{sgx} and \gls{tdx}~\cite{costan2016sgx,intel2023tdx}, AMD \gls{sev}~\cite{amd2020sevsnp}, and ARM \gls{tz}~\cite{pinto2019trustzone} and \gls{cca}~\cite{arm2022cca} are now deployed across data centers, edge devices, and consumer hardware alike.
\gls{tee} remote attestation is used in applications including privacy-preserving machine learning and others~\cite{andreoletti2026llm,hazyresearch2025trustgap,li2025teeslice}.

However, directly transplanting terrestrial \gls{tee} architectures to orbital platforms is insufficient, for two reasons.
First, existing \glspl{tee} were designed under a threat model that \emph{assumes} physical access is possible and attempts to defend against it — yet their defenses are increasingly failing.
Process-based \glspl{tee} such as \gls{sgx} have been shown vulnerable to a growing number of side-channel and microarchitectural attacks~\cite{batteringramsp26,tee-fail}.
\gls{vm}-based approaches like \gls{sev} and \gls{tdx} trade enclave-level granularity for a substantially larger \gls{tcb} that includes the entire guest kernel, expanding the attack surface in a different dimension, while remaining susceptible to side-channel attacks from the hypervisor~\cite{heraclesCPA2025,wiretap}.
More fundamentally, all existing \gls{tee} architectures share a critical limitation: they cannot defend against an adversary with sustained physical access to the host machine.
An attacker who can probe buses, extract memory contents, or tamper with hardware peripherals can undermine the very root of trust on which these systems depend, as demonstrated by recent attacks~\cite{wiretap,tee-fail}.
These are not implementation bugs but inherent limitations of terrestrial trusted computing.

Second, and for the orbital context, more pressing, current \glspl{tee} root their attestation chains in secrets provisioned during manufacturing: keys that the hardware vendor held, or may continue to hold, creating a pre-launch secret window and a single-vendor trust dependency that the verifier cannot independently audit~\cite{flashbots_ztee2_2024}.
For decentralized satellite networks, where multiple operators must mutually verify each other's platforms without relying on a shared vendor, this model is fundamentally misaligned.
Recent works aim to close the gap between the execution environment and its physical location, suggesting the need for extended attestation capabilities that can bind the workload to its execution location~\cite{rezabek2025dcea}.
In addition, most current \gls{tee} solutions lack verifiability of firmware and hardware, which is part of the \gls{tcb}, and struggle with supply-chain attacks where the private keys used as a root of trust are provisioned during manufacturing.
A trust architecture for orbital infrastructure must therefore be bootstrappable without pre-provisioned secrets, auditable across vendor boundaries, and verifiable without dependence on any single manufacturer's attestation service.

A satellite in orbit, however, enjoys a property that no ground-based system can replicate: \emph{physical inaccessibility}.
Once deployed, its hardware is beyond the reach of any adversary short of a nation-state with anti-satellite capabilities, a threat model fundamentally different from on-Earth deployments.
This transforms the design space: rather than building ever more complex defenses against physical attacks, the orbital environment eliminates the threat vector entirely, allowing the security architecture to focus on software-level isolation, cryptographic binding, and distributed verification.

\begin{figure}
    \centering
    \includegraphics[width=.8\linewidth]{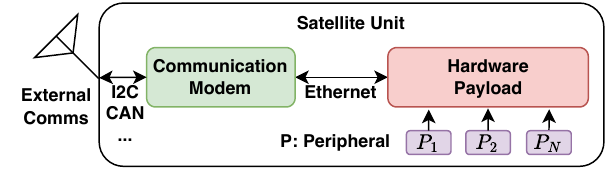}
    \caption{Simplified architecture of a satellite. Contains external communication, which then forwards local signals to the communication modem. The modem is then connected via Ethernet to the hardware payload, e.g., Space Fabric. The payload can include several external peripherals, such as storage devices.}
    \label{fig:satArch}
\end{figure}

This observation motivates \textbf{Space Fabric}, a satellite-native trust architecture that exploits these properties to provide the missing trust foundation for decentralized orbital computing.
Space Fabric relocates the entire trusted computing stack, including \gls{tpm}, \gls{tee}, and execution verification, to orbital infrastructure.
The on-board \gls{tee} provides additional hardware-level isolation for workloads running in orbit.
These components are bound together through the \gls{seap}, a novel protocol that cryptographically links a workload to a specific satellite by incorporating measurements unique to its orbital context and relying on a challenge-response protocol with on-Earth components.
\gls{seap} is the concrete instantiation of a broader security goal we call \gls{poet}: cryptographic evidence that a specific workload executed on a 
specific orbital platform, not inside \emph{some} TEE. This binding ensures that the computation is verifiably executing on the intended satellite, not on a cloned or replicated platform on the ground.
Earth-side components facilitate secure communication with the satellite and allow remote parties to request, submit, and verify workloads, but the trust-critical execution remains entirely in orbit.
\Cref{fig:satArch} outlines the general architecture of the satellite unit, where the hardware payload, e.g., Space Fabric HW, is connected via a communication modem, which provides external communication.

A unique contribution of Space Fabric is fully on-orbit key genesis: all cryptographic signing keys, both satellite identity and attestation keys, are generated within the \glspl{se} after launch, with no secrets existing on Earth at any point.
This eliminates the pre-launch secret window that \glspl{tee} require.
Rather than trusting a vendor to erase provisioned secrets, the verifier's pre-launch trust anchor consists solely of public, non-secret artifacts such as device serial numbers and configuration hashes.
To further reduce single-vendor dependence, Space Fabric distributes its hardware trust anchor across two independent \glspl{se} from different vendors, one closed-source and certified, the other fully open and auditable, both of which must co-sign attestation evidence.
An adversary would need to simultaneously compromise both vendors' silicon to forge a valid attestation proof.

Space, however, is not without constraints.
Limited power budgets, passive thermal dissipation, radiation exposure, and the launch cost per kilogram to orbit all impose strict requirements on the computational hardware that can be deployed.
The current hardware selected for the implementation of Space Fabric is designed with these realities in mind, employing lightweight cryptographic protocols and modest hardware payloads that fit within the envelope of modern small satellite platforms.
Nevertheless, as more computing shifts to space, our architecture should scale as long as its requirements are met.
We make the following key contributions:
\begin{enumerate}[label={\bfseries KC\arabic*}, leftmargin=0.87cm]
\item A systematic threat model for orbital \gls{tee} deployments that introduces an adversary model spanning satellite systems, ground infrastructure, and communication channels with bounded adaptive corruption capabilities (\Cref{sec:threatmodel});
\item The design and implementation of \gls{poet} through \gls{seap}, which cryptographically binds workload execution to a specific orbiting platform through a Byzantine-tolerant endorsement quorum of distributed ground stations, together with a security analysis establishing both satisfaction arguments and impossibility bounds under a strong adaptive adversary (\Cref{sec:architecture} and \Cref{sec:analysis});
\item A fully on-orbit key genesis model in which all cryptographic secrets are generated within \glspl{se} after launch, eliminating the pre-launch secret window present in popular \gls{tee} platforms, complemented by a dual-SE cross-verification architecture that removes single-vendor trust dependence from the attestation chain (\Cref{subsec:root-trust} and \Cref{sec:implementation});
\item An end-to-end implementation of Space Fabric on a USB Armory Mk II with ARM TrustZone, NXP SE050 and TROPIC01 secure elements, and Veraison-based attestation verification, with an evaluation demonstrating mitigation of a broad range of software- and hardware-level attacks (\Cref{sec:implementation}).
\end{enumerate}

% The remainder of this paper is organized as follows. \Cref{sec:background} provides background on existing \gls{tee} architectures, \glspl{tpm}, and satellite computing. \Cref{sec:related-work} surveys related work. \Cref{sec:problemmodel} presents our deployment model and threat model.
% \Cref{sec:architecture} details the Space Fabric architecture and the \gls{seap}.
% \Cref{sec:analysis} provides the security analysis. \Cref{sec:implementation} describes our implementation and argues that it satisfies the security goals.

\section{Background}
\label{sec:background}
This section provides background on the fundamental technologies underlying Space Fabric. We begin by examining \glspl{tee}, comparing process-based and VM-based isolation approaches. 
We then survey the software- and hardware-level vulnerabilities that motivate our design. 
Finally, we discuss \glspl{tpm}, \glspl{hsm}, and satellite computing, and outline Space Fabric's strategy for closing the remaining trust gaps.

\subsection{Trusted Execution Environments}

\glspl{tee} provide isolated execution environments that protect sensitive computations from potentially malicious privileged software. 
Modern \gls{tee} architectures fall into two broad categories: process-based isolation, which protects individual applications or code regions, and VM-based isolation, which secures entire virtual machines.

\subsubsection{Process-Based TEEs}

Process-based \glspl{tee} isolate individual applications or secure enclaves within a running system, protecting them from the operating system and other software components.

\paragraph{Intel SGX} provides hardware-enforced memory isolation through encrypted enclaves~\cite{costan2016sgx}. 
Each enclave operates in its own protected address space with memory encryption/decryption handled transparently by the CPU \gls{mee}. 
A \gls{tee} remote attestation is a cryptographic proof, signed by a hardware-rooted key, that a specific piece of software is running unmodified inside a genuine, isolated \gls{tee} on a particular platform.
Intel \gls{sgx} supports attestation, enabling verifiers to confirm the identity and integrity of enclaves. 
However, \gls{sgx}'s security model explicitly excludes physical side-channel attacks from its threat model~\cite{costan2016sgx}. Nevertheless, researchers have demonstrated practical key extraction and data exfiltration from \gls{sgx} enclaves.
Furthermore, Intel \gls{sgx} provides no protection against memory bus attacks, where adversaries with physical access can probe communication between the CPU and DRAM~\cite{tee-fail}.

\begin{figure}
    \centering
    \includegraphics[width=.6\linewidth]{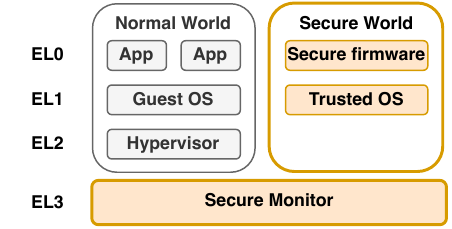}
    \caption{Simplified architecture of an ARM \gls{tz}. \gls{el} ensure separation of individual access layers. }
    \label{fig:arm-tz}
\end{figure}

\paragraph{ARM TrustZone} partitions system resources into two worlds: a secure world for trusted operations and a normal world for untrusted software~\cite{pinto2019trustzone}, highlighted in~\Cref{fig:arm-tz}. 
Unlike Intel \gls{sgx}'s enclave model, TrustZone provides system-wide isolation enforced by the processor's security state. 
The \gls{tzasc} restricts normal-world access to secure memory regions, while \gls{tzpc} governs peripheral access. 
TrustZone does not provide hardware-based remote attestation natively; attestation must be implemented in software, typically through frameworks such as OP-TEE~\cite{suzaki2024opteera}.

\subsubsection{VM-Based TEE Solutions}

VM-based \glspl{tee} extend isolation guarantees to the entire \glspl{vm}, protecting guest \glspl{vm} from potentially malicious hypervisors and host software.

\paragraph{AMD \gls{sev}-\gls{snp}} encrypts guest \gls{vm} memory using per-\gls{vm} keys managed by the \gls{asp}~\cite{amd2020sevsnp}. 
\gls{snp} extends earlier \gls{sev} iterations by adding integrity protection through cryptographic hashes of memory pages, preventing a malicious hypervisor from rolling back or remapping encrypted pages. 
Remote attestation enables verifiers to confirm \gls{vm} identity by measuring the initial memory contents and the CPU state. 

\paragraph{Intel TDX} isolates \glspl{td}—the TDX equivalent of \glspl{cvm}—through hardware-enforced memory encryption and access controls~\cite{intel2023tdx}. 
The \gls{seam} mediates all interactions between the \gls{vmm} and \glspl{td}, enforcing TDX-specific protections and preventing unauthorized memory access. 
\gls{mktme} derives per-\gls{td} encryption keys from hardware secrets, ensuring memory confidentiality and integrity. 
Attestation builds on Intel's \gls{sgx} infrastructure: the \gls{tdqe} generates TD-specific attestation keys, verified against Intel's \gls{pcs}.

\paragraph{ARM \gls{cca}} provides hardware-enforced isolation for confidential workloads on ARM architectures through the \gls{rme}~\cite{arm2022cca}. 
\gls{rme} partitions the physical address space into four distinct worlds: Secure, Non-secure, Realm, and Root, where the Realm world hosts \glspl{cvm} that are protected from both the hypervisor and the Secure world. 
The \gls{rmm} mediates all interactions between the untrusted hypervisor and Realm \glspl{vm}. 
Memory isolation is enforced by the \gls{gpc}, which tags physical memory pages with world assignments and prevents unauthorised cross-world access at the hardware level.

\subsubsection{Software-Level Vulnerabilities}

Despite providing strong isolation boundaries, both \gls{tee} families are susceptible to software-level attacks that exploit the interfaces between trusted and untrusted components.

Process-based \gls{tee} must interact with the untrusted OS for system calls, I/O, and memory management. 
A compromised OS can manipulate enclave inputs and outputs, control page tables, and observe access patterns. 
In SGX, this enables controlled-channel attacks that infer secrets from page-fault sequences, as well as microarchitectural side channels exploiting caches, branch prediction, and speculative execution~\cite{tee-fail,costan2016sgx,batteringramsp26}.

VM-based \glspl{tee} face threats from the hypervisor, which retains control over scheduling, resource allocation, and address translation even though it is excluded from the confidentiality \gls{tcb}. Recent work has demonstrated chosen-plaintext attacks on AMD SEV-SNP via nested page-table manipulation~\cite{heraclesCPA2025}, and the broader challenge remains that the entire guest OS and vendor firmware are included in the \gls{tcb}, substantially expanding the attack surface compared to process-based designs.
%A detailed taxonomy of technology-specific vulnerabilities is provided in Appendix B.

\subsubsection{Hardware-Level Vulnerabilities}
Both \gls{tee} families remain fundamentally vulnerable to physical adversaries. 
Attackers with hardware access can probe memory buses, perform cold-boot attacks to recover DRAM contents, or exploit side channels through power consumption and electromagnetic emanations~\cite{tee-fail,wiretap,abramson2006intel}. TrustZone does not encrypt memory at all, and even state-of-the-art memory encryption has been defeated by recent DDR5 interposition attacks~\cite{tee-fail,wiretap}. 
These are not implementation bugs but inherent limitations of terrestrial trusted computing. These limitations motivate Space Fabric's relocation of trust anchors to physically inaccessible orbital platforms.

\subsection{Trusted Platform Modules}

The \gls{tpm} enhances platform security by providing hardware-rooted measurement, attestation, and key management primitives~\cite{ShepherdTrustcom}. 
Unlike \glspl{tee} which focus on runtime isolation, \glspl{tpm} establish trust in the platform's boot-time configuration and provide cryptographic services bound to that configuration.
During the boot process, each stage is measured and the result extended into \glspl{pcr}, append-only registers where new values are cryptographically combined with existing ones, creating a tamper-evident record of the boot chain. 
The TPM's \gls{ek}, provisioned during manufacturing with a certificate from the \gls{tpm} vendor, anchors the attestation chain~\cite{ShepherdTrustcom}. 
\glspl{ak} derived from the \gls{ek} sign \gls{pcr}-based quotes for remote verifiers without revealing platform identity. 
Sealing operations bind data to specific \gls{pcr} values, ensuring that sealed secrets become accessible only when the platform boots into an expected configuration.
% This complementary approach makes \glspl{tpm} essential for anchoring trust to physical hardware in Space Fabric's architecture.

\subsection{HSM}
A \gls{hsm} is a dedicated, tamper-resistant physical device designed to generate, store, and manage cryptographic keys and perform cryptographic operations, such as signing, encryption, and decryption, within a hardened boundary. 
Its defining property is that private key material never leaves the module in plaintext: applications send data in and receive results back, but the secrets themselves remain locked inside. 
\glspl{hsm} typically enforce access policies, maintain audit logs, and offer protections against both physical probing and logical attacks.
In the context of this work, the NXP SE050 and TROPIC01 fulfill an analogous role at embedded scale: it provides non-exportable key storage, on-chip \gls{ecc} signing, and a \gls{trng} in a package suitable for a satellite board.
While \glspl{tpm} are primarily designed for platform attestation and boot-time measurement, \glspl{hsm} are optimised for general-purpose cryptographic key management and signing, without attestation capabilities and limited programmability.

\subsection{Satellite Computing}
\label{sec:sat-compute}
Satellite computing represents an emerging paradigm where orbital platforms provide computational and communication services beyond traditional ground-based infrastructure. Recent advances in small satellite technology, commercial launch services, and space-qualified computing have enabled deployment of increasingly sophisticated processing capabilities in orbit~\cite{feilden2024starcloud}.

\subsubsection{Physical Security Advantages}
Orbital platforms provide inherent physical security properties unattainable by terrestrial infrastructure.
Physical inaccessibility. Satellites orbiting at altitudes of 500–2000 km \gls{leo}or 35,786 km \gls{geo} remain beyond reach of casual adversaries. 
Accessing a satellite requires either sophisticated ground-based attacks (jamming, spoofing) that target communication links rather than physical hardware, or space operations (rendezvous, docking) that impose extraordinary technical and financial barriers. 
This asymmetry fundamentally alters the threat model: while ground platforms must defend against both software and physical attacks, satellites need to primarily defend against software threats.
Satellites are continuously monitored through ground station networks and space surveillance systems. Any unauthorised approach, physical manipulation, or unexpected orbital manoeuvre would be detected through changes in orbital parameters, communication patterns, or telemetry. While not perfect, sophisticated adversaries might conduct stealthy operations, which raises the bar substantially compared to terrestrial systems, where physical tampering might go unnoticed for extended periods.

Satellite deployment options and latency considerations are detailed in Appendix B.

\subsection{Space Fabric's Strategy}
As computation moves to orbital platforms and decentralized satellite networks emerge, the trust architecture must follow.
The vulnerabilities surveyed in this section reveal two distinct gaps in the current trusted computing landscape. 
The first is the physical access gap: terrestrial \glspl{tee}, regardless of their software isolation guarantees, cannot defend against an adversary with sustained physical proximity to the hardware.
The second is the root-of-trust gap: existing attestation chains depend on secrets provisioned during manufacturing, creating a pre-launch trust window and single-vendor dependency that the verifier cannot independently audit.

Space Fabric addresses both gaps through a layered strategy. 
To close the physical access gap, it relocates the trusted computing stack to orbital infrastructure, exploiting post-launch physical inaccessibility as a passive tamper barrier.
To close the root-of-trust gap, it defers all cryptographic key genesis to orbit, generating identity and attestation keys within co-located \glspl{se} after launch and distributes the hardware trust anchor across two independent vendors, so that no single party's silicon is solely responsible for the attestation guarantee. 
The \gls{seap} then binds these components together, certifying not only \emph{what} program is running inside the \gls{tee}, but also \emph{where}, through a Byzantine-tolerant endorsement quorum of distributed ground stations. 
\Cref{sec:architecture} details this architecture and \Cref{sec:analysis} analyses its security properties and limitations.

\section{Related Work}
\label{sec:related-work}
The trust problem for orbital computing — establishing that a workload executes on a specific satellite, inside a genuine \gls{tee}, with keys that no terrestrial party can forge — has not been addressed by prior work. Existing research spans process-based isolation, \gls{vm}-based confidential computing, lightweight embedded \gls{tee} designs, and distance-bounding protocols, but all operate under the assumption that the hardware resides on Earth. We survey the most relevant threads below and position Space Fabric relative to each.
The \gls{tee} landscape spans process-based isolation, VM-based confidential computing, and lightweight embedded designs.
Ménétrey et al.~\cite{menetrey2022attestation} survey attestation mechanisms across these platforms, noting that TrustZone lacks built-in remote attestation and must rely on software-based solutions such as OP-TEE. 
The IETF RATS architecture \cite{rfc9334} formalises the roles of Attester, Verifier, Relying Party, and Endorser, defining Passport and Background-Check attestation models that \gls{seap}'s exchange instantiates. 
Veraison~\cite{veraison} implements this architecture for heterogeneous platforms. 
Pass, Shi, and Tramèr~\cite{pass2017formal} provide the foundational formal treatment of attested execution, defining an ideal functionality $\mathcal{G}_{\mathsf{att}}$ in the UC framework that captures isolated execution, attestation, and sealing. We use this abstraction for Fabric's combination of GoTEE, SE050, TROPIC01, and SEAP, which is concretely instantiated. 
Google's Confidential Space~\cite{google2023confidentialspace} and Microsoft Azure Attestation~\cite{microsoft2023azure} demonstrate how TEE-based attestation enables multi-party confidential computing in cloud settings; Space Fabric extends this model to orbital infrastructure where physical inaccessibility provides a tamper barrier unavailable in terrestrial data centres. 
Recent work on \gls{dcea}~\cite{rezabek2025dcea} and Intel's \gls{poe}~\cite{intel2024poe} aims to close the gap between where a workload runs and what code it executes, suggesting the need for extended attestation that binds workloads to their physical environments. 
A goal \gls{seap} addresses through its satellite-binding mechanism.

Keystone \cite{lee2020keystone} is an open framework for customisable \glspl{tee} on RISC-V using \gls{pmp} and a small Security Monitor to support remote attestation through SBI calls. 
Sanctum \cite{costan2016sanctum}, an earlier RISC-V enclave design, adds cache side-channel resistance. 
OpenTitan \cite{opentitan} pursues open-source root-of-trust silicon with \gls{puf} based key derivation, though keys are still written to \gls{otp} during a manufacturer-controlled personalisation step, preserving a pre-launch secret window. 
The Tropic Square TROPIC01~\cite{tropic2025tropic01}, used in our architecture, is a fully open and auditable secure element with published RTL, firmware, and a RISC-V core.

WaTZ~\cite{menetrey2022watz} demonstrates that WebAssembly runtimes can execute inside ARM TrustZone with full \gls{ra} support and a formally verified attestation protocol. 
Its small footprint (265 kB) makes it suitable for resource-constrained environments. 
Integrating a similar WASM runtime into Space Fabric's GoTEE Secure World would enable general-purpose, multi-tenant workload deployment through the same \gls{seap} pipeline, which we identify as future work.

As an extension to \gls{seap}, we can also consider distance bounding. Distance-bounding protocols, originating with Brands and Chaum~\cite{brands1994distance}, use tightly timed challenge-response exchanges to upper-bound the physical distance between the prover and verifier, countering relay and distance-fraud attacks. 
Drimer and Murdoch~\cite{drimer2007distance} demonstrated practical distance bounding for smartcard relay attacks, and Avoine et al.~\cite{avoine2021relay} provide a comprehensive survey of the threat model. 
\gls{seap}'s anti-relay mechanism is conceptually related but differs fundamentally: rather than measuring round-trip time at RF speed (impractical over satellite links with variable propagation delays), \gls{seap} exploits the bounded channel corruption window and a Byzantine quorum threshold to ensure at least one honest, uncorrupted \gls{gs} participates in the endorsement. 
The impossibility results in Section 6.3 make explicit that SEAP cannot provide classical distance-bounding guarantees.

As Space Fabric ultimately serves as a security solution to satellites, we consider Falco~\cite{falco2021newspace}, who surveys cybersecurity threats in the New Space era, identifying ground segments and RF links as primary attack surfaces. 
The emerging space edge computing paradigm, exemplified by Thales Alenia Space's Imagin-E payload on the ISS~\cite{chenet2024spaceedge} and ESA's $\Phi$sat-2 CubeSat~\cite{esa2024phisat2} — motivates on-board computational architectures that can be remotely attested and securely updated. 
The Trusted Computing Group's CyRes specification~\cite{tcg2024cyres} addresses hardware roots of trust for resilient satellite platforms, and SpiderOak's OrbitSecure~\cite{spideroak2024orbitsecure} applies zero-trust encryption to satellite data protection. 
To our knowledge, Space Fabric is the first work to combine \gls{tee}-enabled remote attestation with a formally defined satellite-binding protocol and on-orbit key genesis, addressing the trust requirements of decentralized orbital computing where no prior architecture has operated.

\section{System Deployment \& Threat Model} 
\label{sec:problemmodel}

The software- and hardware-level vulnerabilities together with the root-of-trust and supply-chain limitations discussed in~\Cref{sec:background} establish that terrestrial \gls{tee} deployments face risks that cannot be fully mitigated by hardware or software improvements alone. 
%In this section, we cover the system deployments and detailed threat model with involved parties. 

\subsection{System Deployment}
\label{sec:trust-assumption}
Accordingly, the primary objective of Space Fabric is to enable verifiable execution of software on orbital infrastructure that the workload owner does not control. 
The on-board \gls{tee} hosts arbitrary applications, whether operator-managed services or third-party workloads deployed by external tenants onto infrastructure they have no physical or administrative access to. 
In both cases, the core requirement is the same: a verifier that directly interacts with the \gls{ta} must be able to verify what software is running and that it executes within an attested, isolated environment on the claimed orbital platform. 
Space Fabric addresses both scenarios through a single attestation mechanism: cryptographic evidence rooted in on-board hardware trust anchors, endorsed by a distributed \gls{gs} quorum, that proves the identity and integrity of the deployed software, the isolation of the execution environment, and its binding to a specific satellite. 
These guarantees must hold even if the satellite operator behaves adversarially and individual ground stations are compromised, as long as a threshold of honest ground stations participates in the endorsement protocol.

% ....

The Space Fabric strategy for tackling this goal consists of three principal ingredients:
\begin{description}
    \item[TA attestation] The guarantee that a particular \gls{ta} is deployed inside a \gls{tee} is addressed with a hardware-reinforced \textbf{secure bootstrap} procedure binding \gls{ta} to a \gls{tee}. 
    \item[RA attestation] The guarantee that the \gls{tee} executes the expected software stack is addressed via \gls{ra}. A verifier confirms both the integrity of the running software and the authenticity of the hardware producing the evidence. 
    \item[Proof of Execution Triangulation] The guarantee that the \gls{tee} and \gls{ra} attestations come from a satellite is addressed via a \gls{poet}. \gls{poet} is cryptographic evidence that a specific workload executed on a specific hardware satellite platform. \glspl{gs} collectively attest to communicating directly with the satellite using publicly identifiable signing keys.
\end{description} 

\gls{poet} is the overarching security goal: cryptographic evidence that a specific workload executed on a specific orbital platform. 
\gls{seap}, detailed in \Cref{sec:satellite-protocol}, is the concrete protocol that achieves this goal by orchestrating the challenge-response exchange between the satellite's \gls{tee} and the distributed \gls{gs} quorum. In other words, \gls{poet} defines \emph{what} the verifier needs to be convinced of, while \gls{seap} defines \emph{how} that conviction is established.

We model the problem as a game between three main parties, a satellite $\satellite$\footnote{A separate work will address a world with multiple satellites and establish coordination protocols among them.}, ground stations $\gs$, and a tenant $\tenant$ - the party that deploys workloads onto, or relies on attestation evidence from, the orbital platform without trusting its operator. The following assumptions are made about the parties and the communication among them:

\paragraph{Assumptions about satellites.}

A satellite has limited computation capabilities, as outlined in \Cref{sec:sat-compute}. 
Nevertheless, it can perform arbitrary computations. 
To meet Space Fabric's requirements, each satellite carries a \gls{tpm} and a \gls{tee}. 
We denote the \gls{tee} of satellite as $\satellite_\tee$. 
As a feature of \gls{tee}, it can run arbitrary operations and has access to an encrypted storage that is mounted to the \gls{tee} during runtime. 
As used in current deployments on Earth, we assume that the storage's encryption key is given to the \gls{tee} directly via a secure communication channel or can be read from via a driver from e.g., \gls{tpm} or \gls{hsm}. 

Satellites are geographically restricted, and at any given time, a satellite can communicate with a limited set of ground stations (or other satellites) along its trajectory.
The trajectory of every satellite and its geographical location are publicly available~\cite{space-track}. 
These restrictions determine the characteristics of communication between $\satellite$ and $\gsSet$, as detailed below.

\paragraph{Assumptions about ground-stations.}

\glspl{gs} are the on-Earth components for communication. 
There exists a set of \glspl{gs} $\gsSet$. 
For simplicity, a \gls{gs} can be the physical station or a compute unit in its vicinity that delivers data for communication with a satellite. 
\glspl{gs} are geographically distributed, and at any point in time, each \gls{gs} is located within a specific position.
\gls{gs} can be static or dynamic. 
Static corresponds to a large disk satellite, and dynamic can, e.g., be a terminal. 
Depending on that, the position of each \gls{gs} is known or unknown to both satellites and other \glspl{gs}. 
We assume that all \glspl{gs} have access to a global, synchronized clock, e.g., via \gls{ntp}. 

\paragraph{Assumptions about satellite-groundstation communication.}

An honest \gls{gs} communicates with the satellite (only) via a direct downlink/uplink part of the time, e.g., during a window of 10 minutes out of every 90-minute period (an ``orbit'' period).
%\footnote{Future deployments may consider indirect communication channels, such as sattelite-sattelite p2p communication and LEO infrastructure such as Iridium.}
As mentioned above, based on a geographically known schedule of the satellite orbit, $\gs$ can predict windows of time in which direct communication between a $\gs$ and $\satellite$ is possible. 

\paragraph{Assumptions about tenant ground station communication.}

Tenant does not have direct communication with the satellite, and relies on \glspl{gs} to mediate all communication with $\satellite$. 
It deploys workloads into the enclave and communicates with the satellite exclusively through the \glspl{gs}.

\paragraph{Assumptions about identities.}
\label{subsec:global-setup}

\paragraph{Identifiers:} Both satellites and  \glspl{gs} have unique identifiers, including newly joining parties,
which are publicly known by all other parties.
When a communication channel is established between a \gls{gs} and a satellite, each party can retrieve the other's unique identifier, such as the \gls{svn}. 

\paragraph{\gls{pki}:}  We assume the existence of a \gls{pki} to which both satellites and \glspl{gs} have access. 
Specifically, each \gls{gs} $\gs$ has its public key $\pubkey_\gs$. 
This key is relevant for signing, such that $\gs$ can sign messages using the corresponding private key $\privkey_\gs$.
The key $\pubkey_\gs$ of node $\gs$ is known to all \glspl{gs} in $\gsSet$ and all satellites in $\satelliteSet$.
We assume that satellites do not implement \gls{pki}.

\subsection{Threat Model} \label{sec:threatmodel}

We now explore in greater depth the entities that control HW and SW provisioning for satellites and ground stations.
By inspecting all internal hardware components and the bootstrap procedure, we gain insights into the adversarial threats that impact the security of various attestations. 

\subsubsection{Setting \& Trust Assumption}
\Cref{fig:threatmodel-setting} depicts the deployment architecture and entities involved in platform launching in mode details:

\begin{figure}
    \centering
    \includegraphics[width=.9\linewidth]{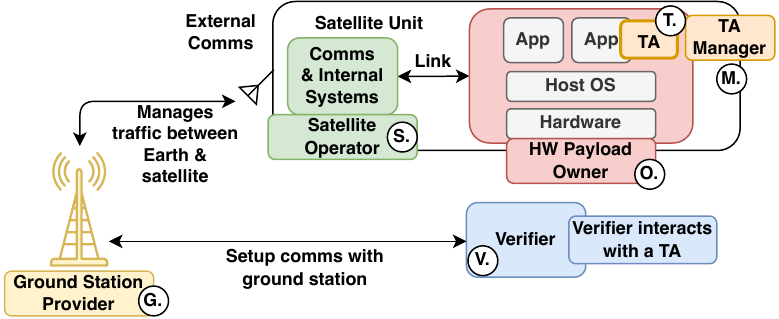}
    \caption{Overview of parties in the threat model. \textbf{Each party's color is used in subsequent figures as a line color}. Depending on the scenarios, some parties may overlap in their roles.
    }
    \label{fig:threatmodel-setting}
\end{figure}

\begin{description}
  \item[Verifier \encircled{V}] anyone who interacts with a \gls{ta} running on the satellite and independently obtains attestation evidence directly from the \gls{tee} to confirm what software they interact with, without having deployed it and without needing to trust the Tenant.
  \item[Ground Station Provider \encircled{G}] operates the ground station infrastructure. 
  The Ground Station Provider establishes a communication channel with the satellite and forwards traffic between terrestrial clients and the orbital platform.
  \item[Satellite Operator \encircled{S}] who operates and manages the satellite unit, including orbital control and communication scheduling (may coincide with \encircled{O} in some configurations).
  \item[HW Payload Owner \encircled{O}] the entity that owns and provisions the physical hardware payload aboard the satellite, including the \gls{tee}, on-board compute, and any embedded \gls{tpm} or root-of-trust component.
  \item[\gls{ta} Manager \encircled{M}] is responsible for managing the lifecycle of \glspl{ta} running within the satellite's trusted execution environment. The TA Manager orchestrates the measured launch of trusted applications on the payload. In some deployments \encircled{M} coincides with \encircled{T} (a tenant deploying and managing their own workload); in others \encircled{M} is the payload owner \encircled{O} or satellite operator \encircled{S}, and \encircled{T} is an external party that relies on attestation to verify what \encircled{M} deployed. 
\end{description}

The HW Payload Owner \encircled{O} is trusted to provision genuine hardware and to maintain physical integrity of the payload prior to launch. 
It is also assumed that the payload has not been physically tampered with post-deployment, and no party engages in attacks that require direct physical access to the orbital hardware.
In particular, the measured launch process for the on-board \gls{tee} and its certificate-issuance infrastructure are assumed to be verifiable. 
The Space Fabric Infrastructure is a composite of the HW Payload Owner \encircled{O}, the Satellite Operator \encircled{S}, and the Ground Station Provider \encircled{G}. 
All software and communication software that make up these components constitute the satellite infrastructure and are considered potentially adversarial, requiring at least a threshold number of \glspl{gs} to be honest. 
These include communication SW controlled by \encircled{S}, OS software and utilities provided by \encircled{O}, and the entire SW stack on the ground provided by \encircled{G}.
Therefore, these systems are \emph{not} trusted to maintain confidentiality for data traversing the communication links or processed in software-visible components on the \gls{gs}, such as the host OS and application layer, which are fully accessible to the \gls{gs} operator; and integrity guarantees are considered meaningful only when they are cryptographically verifiable. 
It is achieved, for example, through measured boot flows on the satellite payload, sealed keys, and hardware-generated attestations relayed over the communication channel.
The primary objective is to provide assurance to the tenant \encircled{T} and verifier \encircled{V} that the deployed workload executes within a trusted on-board \gls{tee} whose hardware integrity was verified prior to launch and remains uncompromised in orbit --- regardless of whether \encircled{T} deployed the workload themselves or is an external party \encircled{V} verifying software managed by \encircled{M}.

\subsubsection{Adversary Capabilities}

\paragraph{Satellite Compromise.}
The adversary may control the satellite's entire software stack outside $\satellite_\tee$, including the host OS, all processing units, and the host-TEE I/O path — intercepting all data transferred to and from $\satellite_\tee$. However, $\satellite_\tee$ itself is assumed secure: $\adv$ cannot launch side-channel or physical attacks.

\paragraph{Ground Infrastructure and Node Corruption.}
The adversary controls the ground station host OS, application layer, and communications stack, and may collude with $\encircled{S}$ or malicious insiders. Additionally, $\adv$ can adaptively corrupt up to $\adversarialGS = \lfloor \frac{\adversarialGSPercentage}{100} \cdot \totalGS \rfloor$ of the $\totalGS$ Earth-based nodes during \gls{seap}'s 
execution, gaining complete control over corrupted nodes' private state and actions. Corruption is irreversible.

\paragraph{TEE Access.}
The adversary controls an arbitrary number of \glspl{tee} on Earth and their communication channels. On-Earth communication between adversarial nodes and \glspl{tee} is assumed instantaneous: if $\adv$ intercepts a message destined for a satellite, it can relay it to a local \gls{tee} and receive a response without measurable delay.

\paragraph{Communication Channel Attacks.}
$\adv$ can act as a \gls{mitm} on any \gls{gs}'s communication channel — intercepting, delaying, reordering, dropping, replaying, or injecting messages — without accessing the \gls{gs}'s private signing key. We impose three restrictions: (i) channel corruption persists for at least $\channelCorruptionWindow$ seconds (rolling windows); (ii) at most $\adversarialChannels$ channels can be corrupted simultaneously; and (iii) if both a \gls{gs} and its channel to $\satellite$ are uncorrupted, then every message received by the \gls{gs} on behalf of $\satellite$ was computed on the satellite. Intuitively, (i) and (ii) model an adversary that cannot reposition faster than the satellite orbits, while (iii) ensures that relaying a message to Earth, computing a response, and returning it to the satellite cannot be done within the communication window.

\section{Space Fabric Architecture}
\label{sec:architecture}
The main technical objective of Space Fabric is to facilitate the certified provisioning and execution of trusted software into space. 

The key novel aspect is to certify not only \emph{what} program is running inside a \gls{tee}, a traditional \gls{ra} offering, but also certify \emph{where} (i.e., in space) this \gls{ta} runs. 
As this is a crucial part of the Space Fabric's proposition, we devised a protocol that leverages the space's native capabilities and communication means. 
In addition, certain challenges arise from existing gaps in tee deployment and execution certification.  

\begin{figure}
    \centering
    \includegraphics[width=\linewidth]{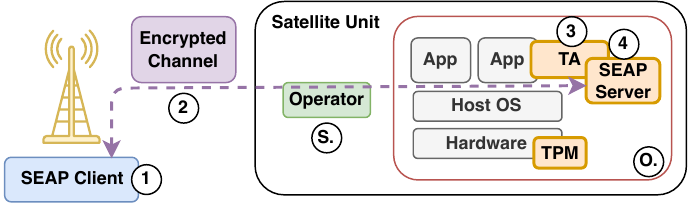}
    \caption{General overview of the Space Fabric Architecture including the \gls{seap} components, and trusted components (in orange) of \gls{ta} and \gls{tpm}.}
    \label{fig:space-fabric}
\end{figure}

To address these challenges, Space Fabric includes three main components, as shown in~\Cref{fig:space-fabric}:

\begin{description}
    \item[Secure-boot] A HW-assisted bootstrap process, proving trusted host-OS deployment on a \gls{tee} or compute node using secure boot, 
    \item[RA] \gls{tee} A remote attestation utility, proving execution on a \gls{tee} based on Arm \gls{tz},
    \item[\gls{seap}] A location certification service, proving \gls{tee} deployment in space, a unique instantiation of \gls{poet}. 
\end{description}

We focus on a single-satellite model, but the approach is generalizable to any satellite as long as it meets the hardware requirements and fits the general setting. 

\Cref{fig:space-fabric} provides an overview of the components and the flow. 
\gls{seap} has two components: the on-Earth verifier/tenant \encircled{1} communicating via an encrypted channel \encircled{2} with the server \encircled{4} running within the \gls{ta} \encircled{3}. We then have \gls{tpm} and \gls{tee} shown in the form \gls{ta}. The \gls{tee} is now visualized as the process-based \gls{tee}, as it is the current implementation we follow, but the \gls{seap} protocol is applicable also to the \gls{cvm}, as long as the requirements and threat model of the \gls{tee} provides the same security guarantees.

\subsection{Root of Trust}
\label{subsec:root-trust}
Space Fabric's root of trust addresses a fundamental limitation shared by existing commercial \gls{tee} platforms: the reliance on secrets provisioned into hardware before deployment. 
Space Fabric eliminates this window by combining three mechanisms. 
The SoC's verified boot facility establishes an immutable first link in the boot chain: the Boot ROM verifies the Trusted OS image against a public root key fused into \gls{otp}, ensuring that only an authenticated software stack executes in the Secure World. 
Crucially, this fused key is a verification key, not a secret.
While a single party holds the image-signing key (unless using, e.g., threshold cryptography), this does not mean a single point of trust because the attestation flow independently measures the running software, and the verifier appraises those measurements against reference values established before launch.
So if the relevant software artifacts are reproducible, it is possible to verify the details as part of the attestation flow. 

Second, all cryptographic signing keys, both satellite identity and attestation keys, are generated on-orbit by co-located \glspl{se} on first boot after launch. 
Private keys are flagged as internally generated and non-exportable under hardware-enforced per-object policies.
No secret material exists on Earth at any point; the verifier's pre-launch trust anchor consists solely of public artifacts such as factory serial numbers, device certificates, SoC identifiers, and configuration hashes.
These artifacts are safe to disclose because they bind attestation to a specific physical device without conveying any signing capability. 
Even if an adversary provisions a second, identically configured device on Earth, the on-orbit key genesis ensures that the two devices produce distinct key pairs.
This argument assumes that key slots are verified to be empty prior to launch.
This would result in different public keys than those endorsed by the \gls{gs} quorum during \gls{seap}, as we introduce in \Cref{sec:satellite-protocol}. 
The verifier therefore rejects attestation evidence from the clone, since its public keys do not match those bound to the satellite's certificate of authorization.
A subtler concern is not hardware cloning but premature key generation on the legitimate device: an attacker with pre-launch physical access could power the board and trigger key generation before launch, producing keys the attacker cannot extract but that would occupy the key slots before the intended on-orbit genesis.
However, the PTA's attested boot logic checks that all \gls{se} key slots are empty before generating keys. %; if slots are already populated, the PTA halts and never produces a valid EAT token.
As the PTA binary is measured, the code is enforcing this property.
Besides, we rely on the fuses for differentiation between testing and production mode, and checks using the monotonic counter to assess the state on first boot. 

The measured boot process complements key generation by producing a tamper-evident record of the platform's software state via the \gls{pcr} extension.
The on-orbit key genesis provides the cryptographic foundation on which \gls{seap} and the \gls{tee} \gls{ra} flow are built. 
\Cref{sec:implementation} details the concrete hardware instantiation and additional implementation information.

\subsection{RA of Trusted Execution Environment}

\begin{figure}
    \centering
    \includegraphics[width=\linewidth]{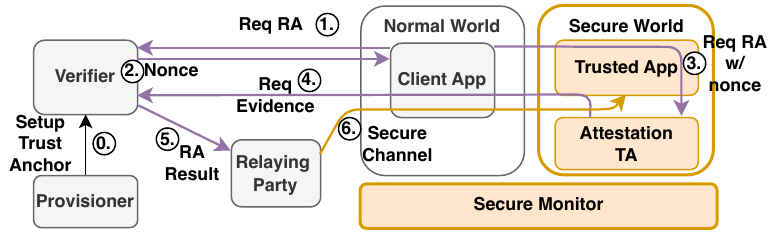}
    \caption{Overview of \gls{ra} for ARM \gls{tz} following OP-TEE approach~\cite{suzaki2024opteera}. The overall goal is to provide a flow for assessing the \gls{ta} and ensure freshness to the request.}
    \label{fig:arm-tz-attest}
\end{figure}

Next, Space Fabric provides remote attestations (\gls{ra}, i.e., signed attestations every time the deployed software produces an externally visible output. 
The current configuration uses ARM \gls{tz}, so SpaceFabric needs to address the lack of native \gls{ra} capabilities.
However, several solutions have been discussed, such as OP-TEE~\cite{suzaki2024opteera,yagawa2024delegating}.
\Cref{fig:arm-tz-attest} highlights the flows using the so-called "Passport Model" known from IETF RATS~\cite{rfc9334}.
In a pre-deployment provisioning step \encircled{0}, the Provisioner loads the attestation public key, signer identity, and reference \gls{ta} hashes into the Verifier. 
At runtime, the client app initiates the attestation \encircled{1} and receives a fresh nonce from the verifier. 
This nonce is forwarded through the Normal World client App into the Secure World, where the \gls{ta} passes it to the Attestation \gls{pta}, a privileged pseudo-TA built into OP-TEE that holds the signing key \encircled{3}.
The \gls{pta} measures the loaded \glspl{ta}, packages the measurements with the nonce into a signed token, and returns this evidence back through the chain \encircled{4}.
Verifier validates the signature against the provisioned trust anchor, checks the nonce freshness, compares \gls{ta} hashes to reference values, and returns an attestation result \encircled{5}.
The critical architectural point is the split between the Attestation \gls{pta} (which owns the secret signing key and runs as part of OP-TEE's kernel) and the \gls{ta} (which is the actual application being attested). 
The \gls{pta} measures the \gls{ta} from a position of higher privilege; it can see the \gls{ta}'s binary hash, but the \gls{ta} cannot access the \gls{pta}'s signing key. 
This separation means the evidence is signed by a component that the \gls{ta} cannot tamper with. 
If attestation succeeds, the Relying Party establishes a secure channel directly with the \gls{ta} \encircled{6}, trusting that it is communicating with verified software running in an isolated \gls{tz} environment.

\subsection{Satellite Execution Assurance Protocol}
\label{sec:satellite-protocol}

Last, Space Fabric needs to certify that remote attestations originate from a genuine space \gls{tee} which is deployed in orbit. 
The challenge is that a \gls{tee} on earth might emulate the space platform design identically. 
It is not easy for a \gls{gs} to verify that it is communicating directly with a satellite. 
The key in Space Fabric is that it would be hard for an adversary to compromise the direct satellite communication links of all ground stations at once. 
Leveraging this point, Space Fabric certifies \gls{tee} location through a distributed consensus protocol among a network of \glspl{gs}.
The protocol is designed around the unique challenges of sporadic, limited communication in space.

\paragraph{TEE setup}
Initially, $\satellite$'s TEE, denoted by $\satellite_\tee$ has access to the
PKI, \ie it has received (as part of its setup) $\pubkey_\gs$ for every
$\gs \in \gsSet$.

\paragraph{Satellite key generation}
The first step that $\satellite_\tee$ does after going online is to create a
new signing key pair, $\langle \pubkey_\satellite, \privkey_\satellite \rangle$.
Note that, since this is generated within $\satellite$'s TEE,
$\privkey_\satellite$ is correctly created and remains private throughout the
execution.

\paragraph{Ground Station ``hello''}
When $\satellite$ is geographically close to an Earth-based \gls{gs} $\gs$, it is possible for $\gs$ to communicate directly with it. To do so, $\gs$ first picks a random nonce $\nonce$ and sends to $\satellite$ a signed ``hello'' message:

$$
\text{hello}, \gs{\nonce, \signature_{\gs, \text{hello}}}
$$

where $\signature_{\gs, \text{hello}} = \sign(\langle \nonce, \text{hello} \rangle, \privkey_\gs)$.
$\gs$ also records the time when the session, identified by $\nonce$, started, \ie when $\gs$ sent the ``hello'' message to $\satellite$.

\paragraph{Satellite ``hello-ack''}
When $\satellite_\tee$ receives an incoming \emph{hello} message by a \gls{gs} $\gs$, it first checks whether $\verify(\pubkey_\gs, \langle \nonce, \text{hello} \rangle, \\ \signature_{\gs, \text{hello}}) = 1$.
If so, then $\satellite_\tee$ responds with a \emph{hello-ack} message as follows:
$$
\text{hello-ack}, \satellite{\nonce, \pubkey_\satellite, \teeProof, \signature_{\satellite, \nonce}}
$$
where $\signature_{\satellite, \nonce} = \sign(\nonce, \privkey_\satellite)$ and $\teeProof$ is a \gls{tee} quote, \ie proof that $\pubkey_\satellite$ was generated inside a \gls{tee}.

\paragraph{Ground Station ``key-verify''}
When $\gs$ receives a \emph{hello-ack} message as above, it checks the following:
(i) if the session identified by $\nonce$ started within $\channelCorruptionWindow$ seconds; (ii) if $\verify(\pubkey_\satellite, \nonce, \signature_{\satellite, \nonce}) = 1$; (iii) if $\teeProof$ is valid.
If all checks hold, then $\gs$ creates the following signature $\signature_{\gs, \timestamp, \satellite} = \sign(\langle \pubkey_\satellite, \timestamp \rangle, \privkey_\gs)$, where $\timestamp$ is the timestamp of $\gs$'s local clock at the point of the signature generation. 
$\gs$ then creates a \emph{key-verify} message, as below, and sends it to $\satellite$:
$$
\text{key-verify}, \gs{\timestamp, \signature_{\gs, \timestamp, \satellite}}
$$

\paragraph{Certificate of identification}
After receiving a \emph{key-verify} message, $\satellite_\tee$ first checks that the signature is valid, \ie whether
$\verify(\pubkey_\gs, \langle \pubkey_\satellite, \timestamp \rangle, \signature_{\gs, \timestamp, \satellite}) = 1$.
If the signature is valid, then $\satellite_\tee$ stores the message and attempts to create a certificate of authorization. 
To do so, $\satellite_\tee$ retrieves $\adversarialGS + 2 \cdot \adversarialChannels + 1$ of the stored \emph{key-verify} messages, where the timestamps of any two messages in this subset differ by no more than $\channelCorruptionWindow$ seconds.

When $\satellite$ has received $\adversarialGS + 2 \cdot \adversarialChannels + 1$ such \emph{key-verify} messages, it concatenates them to create a certificate of authorization $\certificate_\satellite$, which is also signed to prove that the concatenation was done inside the \gls{tee}.

For any subsequent message signed by $\privkey_\satellite$, $\satellite_\tee$
can provide $\certificate_\satellite$ to convince any party that it has successfully completed the identification protocol.

% possibly move to appendix
Appendix~\ref{appendix:pseudocode} introduces algorithm~\ref{alg:satellite-protocol-sat} and algorithm~\ref{alg:satellite-protocol-gs} which are run by $\satellite_\tee$ and $\gs$ respectively.

\subsection{Combining \gls{tee} \gls{ra} with \gls{seap}}
After identifying the \gls{ra} approach for the \gls{tee} and \gls{seap} flows, we must combine them to provide users with confidence that the satellite actually operates in space as part of the attestation flow. 
We consider the \gls{seap} to run only once, until a sufficient number of votes is collected, as part of the setup. 
Nevertheless, the lifecycle of the protocol specifics is outside of the scope of the paper.

\begin{figure}
    \centering
    \includegraphics[width=\linewidth]{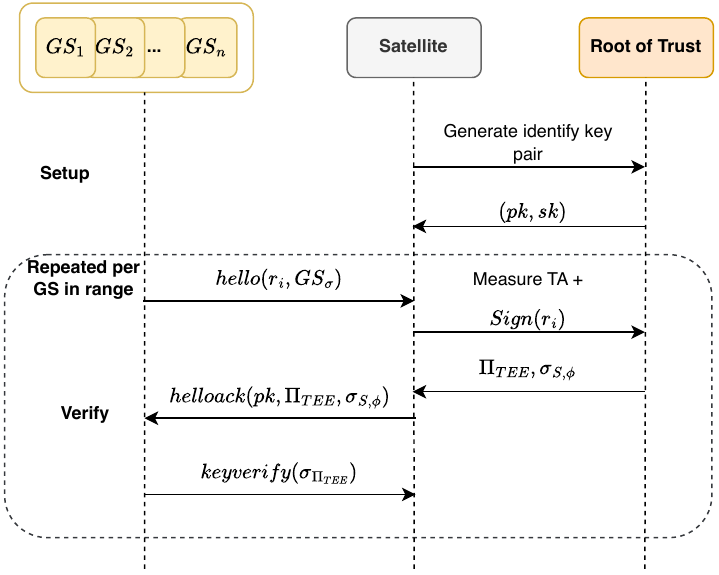}
    \caption{SEAP combined with \gls{tee} \gls{ra}. The satellite's \gls{tee} generates a fresh identity key pair within the Root of Trust. Each \gls{gs} within range executes a three-message exchange. If all checks pass, \gls{gs} endorses the satellite's public key with a timestamped \texttt{key-verify} signature.}
    \label{fig:arm-tz-attest-seap}
\end{figure}

\Cref{fig:arm-tz-attest-seap} covers the sequence of the steps for integration of \gls{seap} with \gls{ra} flow.
During setup, the satellite's \gls{tee} performs a verified boot and generates a fresh identity key pair $⟨\pubkey_S, sk_S⟩$ inside the \gls{rot}, where $sk_S$ is hardware-bound and never exported. 
When a ground station $GS_i$ comes into range, it initiates a three-message \gls{seap} exchange: $GS_i$ sends a signed \emph{hello} with a random nonce $r_i$, which the \gls{tee} verifies against the pre-provisioned ground station \gls{pki}. 
The \gls{tee} then runs an internal attestation sub-protocol - measuring its running software, packaging the claims with $r_i$, and having the \gls{rot} sign the evidence to produce $\pi_{TEE}$.
The \gls{tee} also signs $r_i$ with $sk_S$ and returns everything in a \emph{hello-ack} message: the satellite's public key $\pubkey_S$, the \gls{tee} attestation evidence $\pi_{TEE}$, and the signed nonce $\sigma_S,r_i$.
The ground station performs three checks: (i) the session is within the time window $\channelCorruptionWindow$, (ii) the nonce signature $\sigma_S,r_i$ verifies under $\pubkey_S$, and (iii) $\pi_{TEE}$ appraises successfully via the Verifier, confirming the key was generated in genuine, correctly configured hardware. 
If all checks pass, $GS_i$ endorses $\pubkey_S$ by signing it with a timestamp and returning a \emph{key-verify} message.
The \gls{tee} stores this endorsement. 
As the satellite orbits and contacts successive ground stations, it accumulates endorsements until it holds $\adversarialGS + 2 \cdot \adversarialChannels + 1$  valid ones within the time window — enough to tolerate $\adversarialGS$ adversarial \glspl{gs} and $\adversarialChannels$ adversarial relay channels. 
The \gls{tee} then bundles these endorsements and signs them via the \gls{rot} to produce $Cert_S$, a certificate of authorization that any relying party can verify against the \gls{gs} \gls{pki} without re-running the attestation protocol.

\subsection{Ground Station Committee Update Protocol}
\label{sec:committee-protocol}
As the \gls{seap} plays a crucial role in security, it is important to briefly consider the lifecycle of \gls{gs} on Earth.
The \gls{gs} on-Earth can join and leave the respective committee, and it is crucial that the number of members does not fall under the security threshold of votes collected by the satellite. 
Essentially, the process will result in the public key of a new \gls{gs} being added to the registry of public keys to which the \glspl{gs} and satellites have access.

\paragraph{Handover} We treat the existing set of \glspl{gs} as a committee.
When a new committee $\gsSet'$ needs to be formed, then the \glspl{gs} in $\gsSet$ create a \emph{handover} certificate which establishes the new $\gsSet'$ as the canonical committee.
This certificate includes the public keys of $\gsSet'$ and a signature from at least $\adversarialGS + 1$ parties from $\gsSet$.

\paragraph{Key rotation}
When a new committee is being established, each honest \gls{gs} $\gs$ does the following:
\begin{enumerate}
    \item Creates a fresh public key $\pubkey'_\gs$ and is registered to the new committee with it;
    \item Securely removes the old key $\pubkey_\gs$ after the handover certificate is issued.
\end{enumerate}

Finally, when a handover certificate is created, the honest \glspl{gs} send it to the satellites, who in turn update their local set of committee public keys.

\section{Security Analysis}
\label{sec:analysis}
This section outlines the general goals of the Space Fabric to the user, the properties of the protocol, and several attacks we consider along with their mitigation. 

\subsection{Security Goals}

\label{sec:goals-security}
Motivated by the terrestrial \gls{tee} vulnerabilities surveyed in~\Cref{sec:background}, we design Space Fabric, which covers \gls{tpm}, \gls{seap}, and the use of a \gls{tee} on a satellite. 
Overall, Space Fabric is designed to provide a ground-based (or space-based) verifier with strong, enforceable guarantees about code executing aboard an orbital platform.
At a high level, the verifier must be convinced that attestation evidence originates from a genuine \gls{tee} resident on a specific satellite, is cryptographically bound to that physical platform, and cannot be substituted, replayed, or proxied from a terrestrial or alternative orbital host.

\paragraph{G1 — Authentic Orbital Execution:} The verifier must obtain evidence that the attested workload executes within a \gls{tee} on the claimed satellite and not on any ground-based surrogate or emulated environment. 
Evidence must be unforgeable and rooted in hardware-backed keys provisioned after launch, and the satellite's physical inaccessibility after deployment provides a passive tamper barrier absent in terrestrial deployments.

\paragraph{G2 — Root-of-Trust Integrity:} The signing keys underpinning attestation must be device-unique and must not have been extractable during manufacturing or integration. Overall, it is essential that the keys are generated on the device.

\paragraph{G3 — Platform Binding and Anti-Relay:} Attestation reports must be cryptographically bound to the unique identity of the on-board platform. 
A malicious \gls{gs} or a compromised communication relay must be unable to substitute evidence from another \gls{gs}.
Nonce-binding and freshness mechanisms must remain effective even when all communication channels between the satellite and verifier pass through potentially adversarial ground infrastructure.

\paragraph{G4 — Measurement Freshness and Consistency:} Evidence must reflect the current execution state of the \gls{tee} at the time of the request. 
Stale or pre-recorded attestations must be detectable.

\paragraph{G5 — Auditability Under Supply-Chain Uncertainty:} Because satellite hardware undergoes integration across multiple actors prior to launch, the protocol must not require the verifier to trust any single vendor attestation service. Reference measurements and trust anchors must be established and sealed prior to launch, allowing post-deployment verification to proceed without live interaction with the hardware manufacturer.

We explicitly exclude post-launch physical attacks on the spacecraft bus as outside our threat model, consistent with the assumption that on-orbit hardware is physically inaccessible to adversaries. 
Supply-chain trojans introduced before launch are similarly excluded, though \textbf{G5} limits the degree of trust extended to any single integration party. Together, these goals define the basis for analyzing \gls{seap}'s protocol construction in the following sections.

%\subsection{Ideally TEE functionality}
% Consider defining here this one and argue how do we satisfy it later on even with a weaker TEE.

\subsection{SEAP Properties}
We outline the security rationale of the protocols of \Cref{sec:satellite-protocol} and \Cref{sec:committee-protocol}. 
First, we define the properties that the protocols should aim to guarantee. 
We outline, on a high level, how these properties are achieved.

\subsubsection{Satellite Identification}

The satellite joining protocol of \Cref{sec:satellite-protocol} should aim to satisfy two main properties, \emph{satellite availability} (\Cref{def:availability}) and \emph{correctness} (\Cref{def:correctness}). 
Availability ensures that all satellite-based \glspl{tee} eventually join the system, whereas correctness guarantees that no Earth-based \glspl{tee} join. 

\begin{definition}[Satellite Availability]\label{def:availability}
    Let $\satellite_\tee$ be a \gls{tee} hosted on a satellite $\satellite$.
    Availability is guaranteed, with parameter $\availabilityParam$, if $\satellite_\tee$ successfully completes the identification protocol of \Cref{sec:satellite-protocol} and produces a certificate of authorization at most $\availabilityParam$ seconds after $\satellite$'s stabilization.
\end{definition}

\begin{definition}[Correctness]\label{def:correctness}
    Let $\satellite_{\hat{\tee}}$ be a \gls{tee} that is \emph{not} hosted on any satellite.
    Correctness is guaranteed if $\satellite_{\hat{\tee}}$ does not produce a certificate of authorization (unless with negligible
    probability).
\end{definition}

\subsubsection{Committee Update}

We now outline the properties that the committee update protocol should guarantee. 
The first property, \emph{committee availability} (\Cref{def:committee-availability}), posits that the adversary should not be
able to block a committee update. 
Second, the update protocol should guarantee that the adversary cannot control at any point more than $\adversarialGSPercentage$\% of any historical committee's \glspl{gs}; we call this property \emph{resistance to posterior corruptions} (\Cref{def:posterior-corruption-resistance}).

\begin{definition}[Committee Availability]\label{def:committee-availability}
    Let $\gsSet$ be a committee of Earth-based \glspl{gs}. 
    If all honest \glspl{gs} in $\gsSet$ are given as input a new committee $\gsSet'$, then a handover certificate from $\gsSet$ to $\gsSet'$ is created with overwhelming probability.
\end{definition}

\begin{definition}[Resistance to Posterior Corruptions]\label{def:posterior-corruption-resistance}
    Let $\gsSet_0, \dots, \gsSet_k$ be a sequence of committees, such that there exists a correct handover certificate for any pair $(\gsSet_i, \gsSet_{i+1})$, where $i \in [0, k-1]$.
    For every committee $\gsSet_i$, each party $\gs \in \gsSet_i$ be identified by a public key $\pubkey_{\gs, i}$ (with a corresponding private key $\privkey_{\gs, i}$). 
    At every point in time and for every committee $\gsSet_i$, it should be infeasible for the adversary to create an aggregate signature that comprises signatures from more than $\adversarialGSPercentage \cdot |\gsSet_i|$ of the committee's parties.
\end{definition}

\subsection{Security Rationale}
\label{subsec:sec-analysis}
% Possibly move to the implementation

The security rationale of \cref{sec:satellite-protocol}'s protocol is based on the following insights. 

\subsubsection{Satellite Identification}
First, it suffices to prove that at least one honest \gls{gs}, whose channel is not corrupted, attests that a \gls{tee} $\satellite_\tee$ is on a satellite. 
If an honest \gls{gs} $\gs$'s channel is uncorrupted, then, for any messages that $\gs$ receives which are proven to be created by a \gls{tee}, $\gs$ is assured that said \gls{tee} is on a satellite, due to the discussion in \Cref{sec:trust-assumption}.

Second, the \gls{tee} attestation $\teeProof^\pubkey$ proves that the public key $\pubkey_\satellite$ was created inside a \gls{tee} that runs the given protocol. 
This suffices as proof that every signature, which is verifiable by $\pubkey$, was created by a \gls{tee} that runs Algorithm~\ref{alg:satellite-protocol-sat}. 
This guarantees that, first, $\adv$ does not have direct access to $\privkey_\satellite$, \ie cannot create arbitrary signatures. 
Second, it proves that the creation of the certificate $\certificate_\satellite$ is conditional on performing the timestamp window check of Line~\ref{alg-line:time-window-check} (Algorithm~\ref{alg:satellite-protocol-sat}).

Third, at any point in time, an adversarial \gls{tee} which is based on Earth can collect attesting signatures from at most $\adversarialGS + 2 \cdot \adversarialChannels$.
Since $\adversarialGS$ \glspl{gs} are adversarial, meaning that $\adv$ can issue signatures arbitrarily, the main argument here concerns channel corruptions, as follows.
The protocol requires that each session between a \gls{gs} and a satellite, which is identified by a nonce $\nonce$, lasts at most
$\channelCorruptionWindow$ seconds (cf. Algorithm~\ref{alg:satellite-protocol-gs}). 
Additionally, by assumption, each channel corruption lasts at least $\channelCorruptionWindow$ seconds.
Therefore, at any point in time, an adversarial \gls{tee} can have at most $\adversarialChannels$ active sessions with different \glspl{gs}.

The first observation is that \gls{tee} does not know when a channel corruption starts. 
The only information that \gls{tee} has is the timestamp (signed by \gls{gs}), \ie the timestamp of the interaction with \gls{gs}. 
Therefore, \gls{tee} can check whether the received timestamps, \ie the interactions with \glspl{gs}, occurred within a certain time window, which depends on $\channelCorruptionWindow$.

The second observation is that communication between a \gls{gs} and a \gls{gs} over a corrupted channel lasts less than $\channelCorruptionWindow$ seconds.
Therefore, it is possible that, within some window of $\channelCorruptionWindow$ seconds, the \gls{tee} communicates with two \glspl{gs} whose channels were corrupted within more than $\channelCorruptionWindow$ seconds.
For example, let $\channelCorruptionWindow = 6$ seconds and $\adversarialChannels = 1$. 
Now, imagine that \gls{gs} $\gs_1$'s channel is corrupted at time $t_0$ and $\gs_1$ completes the protocol with a \gls{tee} at time
$t_0 + 3$. 
Next, at time $t_0 + 6$ $\adv$ releases $\gs_1$'s channel and corrupts the channel of \gls{gs} $\gs_2$. 
Following, $\gs_2$ completes the protocol with the \gls{tee} at time $t_0 + 7$. 
Observe that: (i) the adversarial channel corruption assumption is satisfied (since no two channels were corrupted at the same time) and (ii) the \gls{tee} completes the protocol with both \glspl{gs} within $4 < \channelCorruptionWindow$ seconds.

By extending the attack to an arbitrary number of corrupted channels, we conclude that an adversarial \gls{tee} can obtain signatures from at most $2 \cdot \adversarialChannels$ \glspl{gs} with corrupted channels.

\emph{Note:} One might be tempted to resolve the attack by setting the signature threshold at $\adversarialChannels$ and decreasing the timestamp window check (Line~\ref{alg-line:time-window-check} of Algorithm~\ref{alg:satellite-protocol-sat}).
However, this line of defense would not work, because $\adv$ controls message delivery between the \gls{gs} and the \gls{tee}, so it can control the timestamp of the \gls{gs}'s messages.
In the example of the above paragraph, assume that the \gls{tee} checks whether any two interactions happened within $\frac{channelCorruptionWindow}{2} = 3$ seconds.
In that case, $\adv$ can ensure that $\gs_1$'s interaction happens on time $t_0+5$ and
$\gs_2$'s interaction happens at time $t_0+7$, which would again bypass the check.
Essentially, $\adv$ can perform the first \gls{gs}' interaction arbitrarily
close to the end of the first corruption window and the second \gls{gs}' interaction arbitrarily close to the beginning of the second corruption window, such that the time difference between them is arbitrarily small.

In conclusion, by requiring that the satellite's certificate of authorization include signatures from $\adversarialGS + 2 \cdot \adversarialChannels + 1$, we ensure that at least one honest \gls{gs} performed the identification protocol with the satellite's \gls{tee} over an uncorrupted channel.

\subsubsection{Committee Update}
Rotating committees is a well-known and solved problem in cryptographic literature. 
Since all committee members are identifiable by public keys, creating a certificate signed by at least one honest party guarantees that the new committee is correct.

Regarding availability (\Cref{def:availability}), we remind that the adversary $\adv$ controls a percentage $\adversarialGSPercentage$\% out of all $\totalGS$ \glspl{gs}.
Since the certificate must be signed by at least one honest party, the signers must be at least $\adversarialGS + 1$. 
If a majority of \glspl{gs} is adversarial, that is, if $\adversarialGSPercentage > 50 \Leftrightarrow \totalGS \leq 2 \cdot \adversarialGS + 1$, then the number of honest \glspl{gs} is smaller than $\adversarialGS + 1$. 
In other words, a certificate would need to be signed by at least one adversarial party, which, in turn, means the adversary can block all certificate issuance, thereby violating availability. 
Therefore, to guarantee availability, it should hold that $\adversarialGSPercentage < 50 \Leftrightarrow \totalGS \geq 2 \cdot \adversarialGS + 1$, \ie, the majority of \glspl{gs} is honest. 

Regarding posterior corruptions, let us first provide intuition as to why this property is needed.
Consider the following scenario. 
The first committee consists of $10$ parties, none of which are corrupted. 
After the first committee changes, the second committee consists of $30$ parties.
At this point, $\adv$ corrupts the $10$ parties of the first committee; note that the honest majority requirement (needed for availability) is still satisfied for the latest committee. 
Afterwards, a satellite $\satellite$ whose \gls{tee}, upon deployment, holds the ``genesis'' public key set of the first committee, joins the network. 
Since $\adv$ control all communication of $\satellite_\tee$, $\adv$ can prevent $\tee_\satellite$ from obtaining the handover certificate, so $\satellite_\tee$ can complete the protocol of \Cref{sec:satellite-protocol} by only communicating with corrupted \glspl{gs} and, essentially, never join the system.

To avoid this hazard, we observe that the root cause of it is posterior corruptions, \ie the ability of $\adv$ to corrupt a \gls{gs} $\gs$ and use the key that $\gs$ used in all past committees. 
If resistance to posterior corruptions (cf. \cref{def:posterior-corruption-resistance}) is guaranteed, then $\adv$ cannot produce signatures for the keys of old committees. 
In our protocol, this is ensured by the key-deletion step after each handover committee is created. By evolving its key, even if $\gs$ is corrupted later, $\adv$ cannot create signatures on behalf of $\gs$ for past committees.

Essentially, this property ensures that the keys of all committees satisfy the adversarial corruption bound at all times. Therefore, as long as one committee authorizes a new satellite $\satellite$'s key, $\satellite$ does not have to get re-authorized with every committee change.

\subsection{Challenges}
\label{sec:challenges}
We now present a set of challenges that delineate the limits of the goals our protocol can achieve.

\paragraph{\textbf{C1:} Adversarial Satellite Control}

As described in \cref{sec:threatmodel}, $\adv$ controls the entire satellite except $\satellite_\tee$. 
In this setting, \emph{availability} cannot be guaranteed. 
Because $\adv$ controls all communications channels of $\satellite_\tee$, it can completely block all communication from/to $\satellite_\tee$ and isolate it from every other party. 
% Therefore, the goal of our identification protocol will be
% to guarantee correctness, meaning that only satellite-based TEEs are
% authorized, as opposed to liveness (guaranteeing that all satellite-based TEEs
% are authorized), which is impossible under such strict assumptions.

\paragraph{\textbf{C2:} Adversarial Message Relay}
\label{subsec:impossibility-relay}

\begin{figure}
    \begin{center}
        \includegraphics[width=0.8\columnwidth]{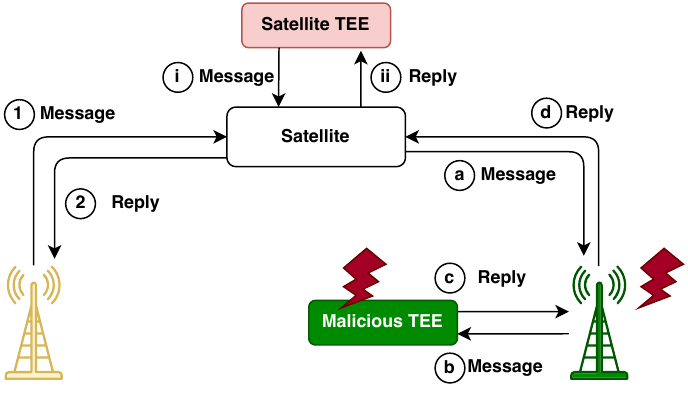}
    \end{center}
    \caption{
        Illustration of a relay attack. In the first scenario (messages 1-i-ii-2), the computation is performed (correctly) in the satellite-based \gls{tee}.
        In the second scenario (messages 1-a-b-c-d-2), the computation is done (adversarially) in an Earth-based \gls{tee}.
        In our identification protocol, both scenarios are indistinguishable from the \gls{gs} 's perspective.
    }
    \label{fig:relay-attack}
\end{figure}

In \Cref{sec:threatmodel}, we required that if a \gls{gs} $\gs$ and its communication channel with a satellite $\satellite$ are uncorrupted, then messages sent to $\gs$ by $\satellite$ (if any) are computed \emph{on the satellite}.
This requirement is necessary, since \emph{correctness} cannot be guaranteed otherwise.

We illustrate this impossibility in \Cref{fig:relay-attack}.
In detail, we consider a setting where an Earth-based \gls{gs} sends a message to a \gls{tee} and receives a reply. 
Here, there exist two scenarios. 
In the first scenario, outlined by the messages 1-i-ii-2, the computation of the reply occurs within a \gls{tee} that is based on the satellite. 
This is the honest scenario, where the satellite forwards the messages correctly to its \gls{tee}.

In the second scenario, identified by the messages 1-a-b-c-d-2, the computation occurs in a \gls{tee} stationed in an Earth \gls{gs}.
Here, the satellite relays the \gls{gs}'s message back to an Earth-based station, which then uses a \gls{tee} based on Earth to produce the computation. 
Note that, since we assume that $\adv$ controls all satellite operations except the \gls{tee}, $\adv$ can intercept the original message and relay it as shown.

In the context of our identification protocol, the Earth-based \gls{gs}, which sends the original message, has no information about the satellite's \gls{tee}.
Therefore, the two scenarios shown above are indistinguishable from the point of view of the \gls{gs}. 
For this reason, if the second scenario (relaying to Earth and back) is possible, then the \gls{gs} would authorize the public key produced within the Earth-based \gls{tee}, even if both the \gls{gs} and the communication channel with the satellite are uncorrupted.
Therefore, we consider the combination with on-board \gls{tpm}, which could be registered prior to the launch and the \gls{tee} flow, or the \gls{tee} flow with general keys registered on-Earth before launch in the Veraison with \gls{hsm}. 

\paragraph{\textbf{C3:} Adversarial Channel Control}
\label{subsec:impossibility-channel}

In \Cref{sec:threatmodel}, we required that $\adv$ can corrupt the communication channels of at most $\adversarialChannels$ Earth-based \glspl{gs} at any point in time. 
This assumption is necessary; otherwise, if all channels are corrupted, then neither \emph{availability} nor \emph{correctness} can be guaranteed.

\begin{figure}
    \begin{center}
        \includegraphics[width=0.8\columnwidth]{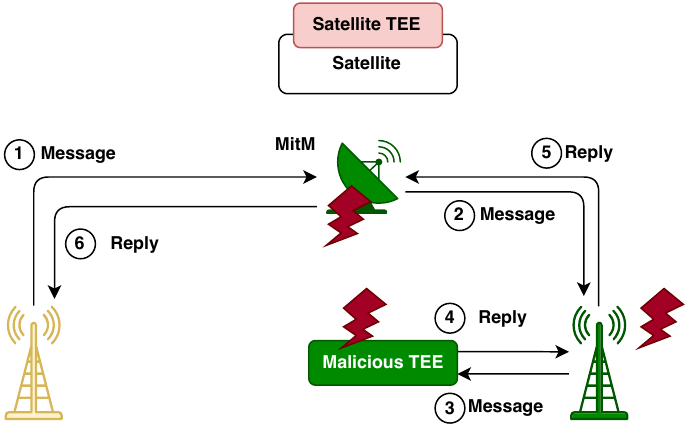}
    \end{center}
    \caption{
        Illustration of a \gls{mitm} attack. The adversary controls the communication channel between the \gls{gs} and the satellite, so it blocks all communication towards the satellite and forwards all messages to an Earth-based \gls{tee} that it controls.
    }
    \label{fig:mitm-attack}
\end{figure}

Essentially, when an Earth-based \gls{gs} $\gs$'s channel is corrupted, $\adv$ can act as a \gls{mitm} and forward all messages to an adversarial Earth-based \gls{gs} without being identified by $\gs$. 
If $\adv$ can do this for all \glspl{gs}, i.e., if no \gls{gs} can communicate with the satellite, then it is impossible to authorize the satellite's \gls{tee} (availability attack).
Additionally, because the attack is non-identifiable by $\gs$, the Earth-based adversarial \gls{tee} may complete the identification protocol and obtain an authorization certificate (correctness hazard). 
The \gls{mitm} attack is illustrated in \Cref{fig:mitm-attack}.

\paragraph{\textbf{C4:} \gls{gs} Location Proof}

A problem that should be discussed is whether and how an Earth-based \gls{gs} $\gs$ can prove to a satellite $\satellite$ its location, i.e., whether $\gs$ is on Earth (as opposed to in space) or in a specific geographic area on Earth.
Recall that some of the \glspl{gs} are adversarial, so a simple signed message by the \gls{gs} is not sufficient proof for $\satellite$, which does not know which \glspl{gs} are corrupted.

In our setting, solving this problem is not possible due to the control that the adversary $\adv$ has on the satellite's system. 
Essentially, $\adv$ controls all data that the satellite's \gls{tee} sends and receives. 
Therefore, as with the availability impossibility discussed above, $\adv$ can block any proofs sent by $\gs$. 
Even if communication with $\gs$ is permitted, though, $\satellite_\tee$ does not have direct access to the satellite's sensors.
Therefore, any information regarding the satellite's geolocation that can be intercepted by $\adv$. 
Since $\satellite_\tee$'s only information about $\gs$ is the public key $\pubkey_\gs$, not having access to any such data makes it impossible for $\satellite$ to validate any claims by $\gs$ about its geographical position.

\section{Space Fabric Implementation}
\label{sec:implementation}
This section outlines the instantiation of Space Fabric and its implementation of \gls{seap} and \gls{tee} with unique \gls{rot}. Next, we revisit the security goals and argue how they are answered.

\subsection{Attestation Flow of \gls{seap} and \gls{ta}}
\begin{figure}
    \centering
    \includegraphics[width=0.8\linewidth]{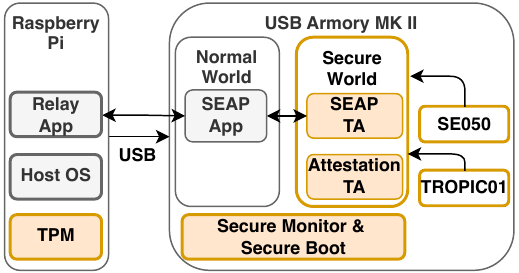}
    \caption{Space Fabric on the satellite platform. The Rpi~5 forwards traffic over USB to the USB Armory Mk~II. The Armory runs the SEAP client app in between the normal and secure worlds. The two \glspl{se} are attached directly to the Secure World to provide key storage and attestation keys.}
    \label{fig:fabricArch}
\end{figure}

We instantiate the \gls{seap} and \gls{ra} protocol on a USB Armory Mk II (NXP i.MX6UL, ARM Cortex-A7) connected via USB to a Raspberry Pi 5 host acting as an untrusted relay~\cite{usbarmory2019mkii}, as shown in~\Cref{fig:fabricArch}. 
The satellite's \gls{tee} is implemented using ARM TrustZone, with GoTEE as the Trusted OS running in Secure World, and a \gls{pta} built into OP-TEE that handles attestation. 
Cryptographic key storage and signing are delegated to the onboard NXP SE050 and TROPIC01 \glspl{se}.
The NXP SE050 is a closed-source, Common Criteria EAL 6+ certified element providing non-exportable \gls{hsm}, on-chip \gls{ecc} signing, a \gls{trng}, and monotonic counters for anti-rollback. 
The TROPIC01, described in~\Cref{sec:background}, provides the open-source counterpart~\cite{tropic2025tropic01}.
The hardware root of trust is anchored in the i.MX6UL's \gls{hab}: the Boot ROM verifies the GoTEE image against a Super Root Key fused into OTP during manufacturing, establishing an immutable first link in the boot chain. 
Endorsements collected during the \gls{seap} exchange are persisted in the eMMC's \gls{rpmb}, which provides authenticated, replay-protected storage. 

We allocate four non-exportable \gls{ecc} slots across two \glspl{se}: SE050 slot~0 holds the satellite identity key $sk_{NXP}$ and slot~1 holds the attestation signing key used to produce $\pi_{TEE_{NXP}}$, and TROPIC01 slot~0 holds the satellite identity key $sk_{TROP}$ and slot~1 holds the attestation signing key used to produce $\pi_{TEE_{TROP}}$.
This dual-SE design serves two purposes. 
First, it strengthens \textbf{G5} (supply-chain auditability) by ensuring that no single vendor's silicon is solely responsible for the attestation guarantee: the SE050 provides a mature, certified, closed-source anchor while the TROPIC01 provides a fully auditable, open-source anchor.
An adversary would need to compromise both vendors (NXP and Tropic Square) simultaneously to forge attestation evidence. 
Second, it enables cross-verification as the \gls{pta} can require both elements to co-sign the \gls{eat} token, producing a dual attestation proof $(\pi_{TEE_{NXP}},\pi_{TEE_{TROP}})$ that the verifier checks against two independent trust anchors.

Under the fully on-orbit model, all four key pairs are generated on-orbit: on first boot, each secure element derives its keys internally. 
The pre-launch trust commitment consists of public, non-secret artifacts from each element: the SE050 factory serial number, the TROPIC01 device certificate (\gls{puf}-bound), the SoC OTP UID, and the hash of each element's locked configuration. 
Under the on-orbit model, Veraison is pre-provisioned with these device identifiers and configuration hashes but not with attestation public keys, since those keys do not yet exist. 
These are bootstrapped during the first \gls{gs} pass via a genesis EAT exchange.
Because the TROPIC01's RTL and firmware are open source, the verifier can independently audit the key-generation and policy-enforcement logic down to the gate level.
When the \gls{pta} executes the \gls{tee} \gls{ra} sub-protocol, it packages the \gls{ta} measurements and nonce $r_i$ into a CBOR-encoded \gls{eat} token and sends it to both \glspl{se} for signing. 
The SE050 (slot 1) produces $\pi^{TEE}_{NXP}$ and the TROPIC01 produces $\pi^{TEE}_{TROP}$.
Similarly, the \gls{pta} signs $r_i$ with both identity keys: $\sigma_{S,r_i}$ via SE050 slot 0 and $\sigma'_{S,r_i}$ via TROPIC01 slot 0. 
The \texttt{hello-ack} carries $r_i, \pubkey_{NXP}, \pubkey_{TROP}, \pi_{TEE_{NXP}}, \pi_{TEE_{TROP}}, \sigma_{S,r_i}, \sigma'_{S,r_i}$ is returned through the relay to $GS_i$.
The \gls{gs} and the verifier independently check both signature chains. 
The Veraison verifier is then pre-provisioned with both attestation public keys, the expected \gls{ta} reference hashes, and the appraisal policy requiring both proofs to be validated.

This architecture also provides a graceful degradation path. 
If one \gls{se} fails (e.g., due to radiation-induced fault or compromised supply chain), the remaining element can continue to produce valid, though single-anchored attestation evidence, which the verifier can accept under a degraded trust policy rather than rendering the satellite completely unattestable.

\begin{figure}
    \centering
    \includegraphics[width=\linewidth]{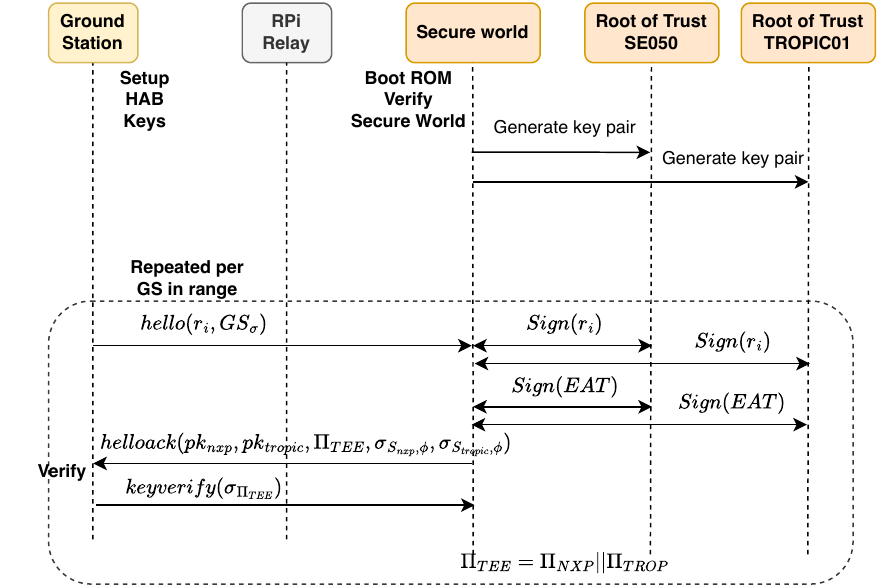}
    \caption{Dual secure element solution with \gls{seap} and \gls{tee} \gls{ra}. The Boot ROM verifies the Secure World image via HAB keys. Both \glspl{se} generate identity key pairs. The \gls{gs} and Veraison verifier independently check both signature chains before issuing a \texttt{key-verify} endorsement. }
    \label{fig:arm-tz-attest-seap-impl}
\end{figure}

\subsubsection{Storage}
We required information on the \glspl{gs} stored in the \gls{tee} to establish the \gls{seap}. An option is to embed the respective public keys in the image, which is verified during boot. 
The \gls{pta} can carry a static trust store — a table of $(GS_i, \pubkey_{GS_i})$ pairs compiled directly into the \gls{ta} or stored as a signed configuration blob loaded alongside it. 
Because the \gls{pta} measures every loaded TA by its SHA-256 hash, any modification to the trust store changes the measurement, which in turn changes the \gls{eat} token the verifier appraises. 
This means the ground station key set is transitively bound to the hardware root of trust: the verifier knows exactly which set of ground station identities the satellite will accept, because that set is reflected in the reference hash it holds.

Besides, during the operation, we envision relying on the storage for usability or state information. 
The \gls{ta} running in Secure World cannot directly access the eMMC block device.
Linux and its storage drivers live in Normal World, and granting the \gls{tee} a raw block-layer path would widen the Secure World attack surface. 
Instead, the typical workflow follows a delegation model. The \gls{pta} issues a Secure World syscall (an RPC request) that crosses the TrustZone boundary and asks the Normal World OS to perform the underlying eMMC I/O on its behalf. 
For ordinary storage, this would be a trust problem, but the \gls{rpmb} frames are authenticated with an HMAC key that only the Secure World holds, so even though the Normal World kernel physically reads and writes the eMMC, it cannot forge, replay, or silently modify \gls{rpmb} contents.

\subsection{Role of the TPM}
The Raspberry Pi 5 relay is untrusted by design — the \gls{seap} protocol assumes the adversary controls everything outside the satellite's \gls{tee}. 
Nevertheless, a discrete TPM 2.0 module attached to the RPi5 provides defence-in-depth by raising the cost of compromising the relay and enabling the verifier to detect tampering that would otherwise be silent.
Besides, many applications can run on the RPi beyond their trusted components running in the secure world. 
The TPM measures the RPi5's boot chain into its \glspl{pcr}. 
Before a \gls{gs} initiates a \gls{seap} exchange, it can request a TPM Quote from the relay, signed by an \gls{ak} certified against the \gls{tpm}'s \gls{ek}.
If the relay's software has been modified, the \gls{pcr} values will not match the reference measurements the \gls{gs} operator published at deployment. 
This does not make the relay trusted in the protocol-theoretic sense, as \gls{seap}'s security does not depend on it, but it provides an operational early-warning mechanism that detects relay compromise.
Such scenarios are also relevant for the communication via the USB to the Armory Mk II.

In addition, the \gls{tpm} provides a natural integration point for post-quantum key material during the pre-launch provisioning phase. 
A PQC key pair (e.g., ML-DSA) can be generated in software on the relay and sealed to the \gls{tpm} under a \gls{pcr} policy reflecting the expected relay configuration. 
The corresponding public key is registered with the \gls{gs} \gls{pki}. 
During \gls{seap} exchanges, the relay can co-sign forwarded messages with the PQC key, providing a quantum-resistant outer signature layer that protects the session against harvest-now-decrypt-later attacks on the classical \gls{ecc} signatures flowing between the satellite and ground station. 
This is generally possible as the \gls{tpm} can serve as an independent root of trust for the \gls{gs}. 

\subsection{Satisfaction of Goals}
We revisit the goals outlined in~\Cref{sec:goals-security} and how our implementation of Space Fabric satisfies them.

\subsubsection{G1 and G2 - Root of Trust \& Authentical Orbital Execution}
\Cref{subsec:root-trust} established that Space Fabric eliminates the pre-launch secret window through on-orbit key genesis and dual-SE cross-verification. 
This section describes how the concrete implementation enforces those properties and compares the resulting trust assumptions against existing platforms, as shown in~\Cref{tab:rot-comparison}.

The SE050 generates both key pairs (slots 0 and 1) via its internal \gls{trng}, with each key object created under the \texttt{ORIGIN\_INTERNAL} and non-exportable policy flags set at creation time. 
The SE050's attested read mechanism allows the verifier to request a signed object attribute report at first contact, confirming that these flags are in force, providing a cryptographic proof that the key policy matches the operator's intended configuration. 
For the TROPIC01, device-unique keys are derived from its PUF, and the fully open-source RTL and firmware allow the verifier to independently audit the key-generation and policy-enforcement logic down to the gate level.

Trust in the on-orbit-generated keys is established during the ground station pass via a genesis EAT token. 
The verifier trusts the genesis EAT because the device identifiers match and the software checks are enforced by the attested application.
The SE050's signed attribute report provides the structural argument: only the chip bearing the pre-registered serial number, operating under the attested policy, could have produced a valid signature. For the TROPIC01, the PUF-bound device certificate serves the same role. Together, these allow the verifier to confirm that both attestation anchors were generated on-orbit without requiring any secret to have existed on Earth.

The trust assumption shifts from the vendor correctly erasing a secret it generated and held to the vendor correctly implementing its published, per-object policy enforcement. 
This is a materially weaker and more auditable form of vendor dependence, further supported by the SE050's Common Criteria EAL 6+ certification and the TROPIC01's fully published design.

\subsubsection{G3 and G4 - Platform Binding \& Freshness}
Each ground station \emph{hello} includes a fresh nonce $r_i$, which the PTA binds into both the \gls{eat} token and a separate signature $\sigma_{S,r_i}$ under the satellite identity key $sk_S$. The \gls{gs} verifies $\sigma_{S,r_i}$ under $\pubkey_S$ and checks session freshness against $\Delta_{corr}$. Because both keys are non-exportable and device-unique, a malicious relay or compromised ground station cannot forge or substitute evidence from another platform — it can only forward messages it cannot re-sign. 

\subsubsection{G5 - Auditability in Supply-Chain}

The implementation relies on many open-source components, such as GoTEE, OP-TEE, and Veraison, allowing any party to audit, reproduce, and independently verify the attestation stack. 
The addition of the Tropic Square TROPIC01 extends this auditability into the \gls{se}.
On the other hand, the NXP SE050 internal design remains proprietary. 
The dual-SE architecture turns this asymmetry into an advantage. 

This open design also enables component substitution: the NXP SE050, the TROPIC01, the i.MX6UL SoC, and the TrustZone-based \gls{tee} are modular choices rather than architectural commitments, so a different \gls{se} or SoC from another manufacturer can be integrated, provided it exposes equivalent primitives (non-exportable key generation, monotonic counters, hardware-verified boot). 
Multiple manufacturers can coexist on the same board, e.g., an NXP SoC paired with an Infineon or Microchip secure element, which would need to collude to break the assumptions.
Still, some components of the HW and SW stacks are not open, so there is space for improvement, but the initial direction is relevant. 

\begin{table*}[t]
\centering
\caption{Comparison of Root-of-Trust and Attestation Key Provisioning across TEE Platforms. Person. - personalization, man. - manufacturing, enf. - enforcement}
\label{tab:rot-comparison}
\resizebox{\textwidth}{!}{%
\begin{tabular}{@{}lllllll@{}}
\toprule
\textbf{Property}
  & \textbf{Intel SGX/TDX}
  & \textbf{AMD SEV-SNP}
  & \textbf{ARM CCA}
  & \textbf{OpenTitan}
  & \textbf{Space Fabric} \\
\midrule

\textbf{Key origin}
  & RPK injected during man.
  & Chip-unique secret fused at production
  & CPAK provisioned by the manufacturer
  & PUF-seeded RootKey written to OTP at man.
 & On-orbit genesis: SE050 via TRNG, TROPIC01 via PUF \\

\addlinespace
\textbf{Pre-launch secret exists?}
  & \textbf{Yes} - Intel holds/held RPK
  & \textbf{Yes} - AMD-held chip secret
  & \textbf{Yes} - manufacturer holds CPAK
  & \textbf{Yes} - Silicon Creator performs person.
  & \textbf{No persistent secret} \\

\addlinespace
\textbf{Vendor trust type}
  & Trust Intel erased RPK copy
  & Trust AMD holds chip secret securely
  & Trust in manufacturer CA and key handling
  & Trust in correct person. + erasure of RootKey
  & Trust in locked slot policies; dual-vendor \\
\addlinespace
\textbf{Single vendor attestation service?}
  & Yes - Intel PCS/IRS
  & Yes - AMD KDS
  & Yes - manufacturer CA
  & No - operator deploys own PKI post-ownership transfer
  & \textbf{No} - dual-SE co-signed \gls{eat}  \\

\addlinespace
\textbf{Design auditability}
  & Fully proprietary
  & Fully proprietary
  & Architecture open; implementation proprietary
  & Fully open-source RTL
  & SE050 closed; TROPIC01 fully opened \\

\addlinespace
\textbf{Hardware binding mechanism}
  & RPK + PPID derived from fused secret
  & VCEK derived from chip-unique fused secret
  & CPAK provisioned into hardware
  & PUF (ring oscillator) entropy + OTP lock
  & SE050 serial + TROPIC01 PUF identity \\

\addlinespace
\textbf{Post-deployment key establishment?}
  & No --- keys pre-exist
  & No --- keys pre-exist
  & No --- keys pre-exist
  & Partial --- Owner Identity at ownership transfer
  & \textbf{Yes} --- dual genesis EAT bootstraps two trust anchors \\

\addlinespace
\textbf{Physical pre-launch attack yield}
  & RPK if Intel erasure claim is false
  & Chip secret if AMD infrastructure compromised
  & CPAK if manufacturer CA compromised
  & RootKey if Silicon Creator person. compromised
  & \textbf{Nothing} --- no secrets to extract \\

\bottomrule
\end{tabular}%
}

\end{table*}

\subsection{Comparison to Existing Attestation Flows}
Our architecture differs from e.g., Intel \gls{sgx}/\gls{tdx} or AMD \gls{sev} in where the root of trust resides, how attestation evidence is produced, and what properties the attestation proves. 
In \gls{sgx}/\gls{tdx} and \gls{sev}, the CPU itself contains a hardware root secret from which attestation keys are derived. 
Both measurement and signing occur entirely inside the processor package, and the resulting evidence is verified through Intel's or AMD's attestation infrastructure.
This allows a verifier to conclude that a specific \gls{td} was launched on genuine manufacturer's hardware, but it requires trust in its key provisioning process and continued availability of attestation services. 
Critically, the current attestations prove \emph{what} code is running but says nothing about where it runs — a gap that recent work on \gls{dcea} and \gls{poe} aims to close for terrestrial deployments.

In Space Fabric, the root of trust is distributed across three components. 
First, the i.MX6UL's \gls{hab} verifies the GoTEE Trusted OS image against a \gls{srk} fused into \gls{otp}, establishing an immutable first link in the boot chain. 
Second, the Attestation \gls{pta} running inside OP-TEE's Secure World measures each loaded \gls{ta}.
Third, two independent \glspl{se}, the NXP SE050 and the Tropic Square TROPIC01, hold non-exportable \glspl{ak} and co-sign the evidence. 
The \gls{pta} communicates directly with both \glspl{se}. 

The resulting dual attestation proof $(\pi^{SE}_{TEE}, \pi^{TS}_{TEE})$ is verified by the operator's Veraison instance against pre-registered public artifacts, anchoring trust in the device hardware and the operator's verification infrastructure rather than a vendor-controlled attestation service. 
An adversary would need to compromise both NXP and Tropic Square simultaneously to forge valid attestation evidence.

Space Fabric also proves \emph{where} it runs through \gls{seap}, which binds the attestation to a specific orbiting platform through a Byzantine-tolerant endorsement quorum that any relying party can verify against the \gls{gs} PKI.
This binding leverages the satellite's post-launch physical inaccessibility as a tamper barrier, closing the physical access gap that terrestrial \glspl{tee}, cannot fully eliminate.
Conversely, \gls{sgx}/\gls{tdx} provides a tighter trust boundary at the silicon level: measurement, key derivation, and signing occur within a single processor package with transparent memory encryption, whereas Space Fabric distributes trust across an SoC and two external \glspl{se} connected over an unencrypted I²C bus. 
Together, these mechanisms provide a hardware-backed attestation architecture for TrustZone systems that approaches the guarantees of processor-integrated \glspl{tee} while remaining independent of vendor-controlled attestation ecosystems, and extends them with orbital binding and cross-vendor auditability that no terrestrial platform currently offers.

\subsection{Applicability to other TEEs}
Although \gls{seap} is instantiated on ARM TrustZone, its core protocol is largely \gls{tee}-agnostic. 
\gls{seap} does not depend on any TrustZone-specific primitive. 
A platform running Intel TDX, AMD \gls{sev}-\gls{snp}, or RISC-V Keystone in a space-grade form factor could participate in the same \gls{seap} exchange, producing equivalent attestation evidence and accumulating endorsements in the same way, provided the \gls{tee} can emit a nonce-bound measurement report. 
The principal challenge in porting \gls{seap} to such platforms is not the protocol itself but the root-of-trust bootstrapping. 
As shown in~\Cref{tab:rot-comparison}, Intel TDX and AMD \gls{sev}-\gls{snp} root their attestation chains in manufacturer-held or manufacturer-provisioned secrets, meaning the vendor's attestation service must remain reachable and trusted throughout the satellite's operational lifetime.
The on-orbit key genesis approach introduced here, anchored in a locked, \gls{hsm} rather than a pre-provisioned vendor secret, is therefore not merely an implementation detail but a necessary adaptation of the trust model to the space environment. 
Future work could explore whether platforms with on-chip \glspl{puf} for key derivation can replicate the same no pre-launch-key property without relying on a discrete \gls{se}.% thereby broadening the set of space-grade hardware on which a full \gls{seap} instantiation is achievable.

\subsection{Addressing the Challenges}

The challenges outlined in \Cref{sec:challenges} outline fundamental protocol limits. 
In practice, the implementation mitigates or bounds each through a combination of hardware binding, physical assumptions, and conservative parameterization.

\textbf{C1:} Availability cannot be guaranteed when the adversary controls all communication outside the \gls{tee}.
We accept this as inherent to the asynchronous Byzantine setting and mitigate it operationally: the \gls{tpm} measured boot chain makes persistent communication-stack tampering detectable, and the \gls{pta} maintains a monotonic heartbeat counter in \gls{rpmb} that the verifier can audit after $Cert_S$ is issued. 
A gap in the expected sequence flags potential censorship, making the attack detectable even if not preventable.

\textbf{C2:} The relay indistinguishability is closed by binding attestation to pre-registered, device-unique hardware identifiers. Under the \gls{tpm} model, the \gls{ekc} is included alongside the \gls{eat} token.
An Earth-based \gls{tee} cannot produce a valid \gls{tpm} Quote under the same \gls{ek}.
Under the fully on-orbit model, the SE050/TROPIC01 factory serial number, SoC OTP UID, and $\mathsf{SlotConfigHash}$ serve the same role: the verifier knows which physical chip must be signing, and replicating it requires possession of the launched hardware. 
The residual trust assumption is that the operator correctly performed device registration with Veraison, a standard operational requirement that does not involve trusting any vendor to erase secrets.

\textbf{C3:} The bounded channel corruption assumption $\adversarialChannels$ is grounded in orbital mechanics. 
For instance, a LEO satellite at approximately 500 km traverses its ground track at roughly 7.5 km/s, with each ground station pass lasting 5–10 minutes. 
Simultaneously compromising RF links at geographically distributed stations requires co-located directional equipment that cannot be repositioned faster than the satellite moves, providing the physical basis for bounding $t_{ch}$. 

$\adversarialChannels$ is set to span the time needed to contact sufficiently many independent stations, typically 6–12 hours for a sparse \gls{gs} deployment.

\textbf{C4:} We accept the impossibility and observe that the security argument does not require the satellite to verify where a \gls{gs} is located, only \textit{who} via $\pubkey_{GS}$ and whether the channel is honest, enforced by the quorum threshold and timing constraints. The geographic distribution of the \gls{gs} infrastructure constrains the adversary's ability to simultaneously corrupt multiple channels, but this is an operational property of the deployment, not an in-protocol guarantee. 
A future hardware extension, e.g., a GPS receiver wired to a \gls{tee} protected peripheral, could allow the \gls{tee} to independently verify its orbital position and reject \texttt{hello} messages outside physically plausible contact windows, at the cost of enlarging the \gls{tcb}.

In summary, the implementation does not resolve the impossibility results cryptographically but bounds them through the combination of pre-registered hardware identities, physical inaccessibility post-launch, and orbital mechanics constraints, assumptions that are explicit, auditable, and strictly weaker than those required by terrestrial \gls{tee} deployments.

\subsection{Protocol Evaluation}
\label{sec:eval}

The \gls{seap} introduces latency and bandwidth costs that must
be assessed against the constraints of orbital platforms. 
Each \gls{seap} exchange with a single ground station comprises three messages
(\texttt{hello}, \texttt{hello-ack}, \texttt{key-verify}), requiring 1.5 round-trip. 
The per-exchange latency is dominated not by propagation delay---which contributes approximately \SI{60}{}-\SI{120}{\milli\second}for a \SI{500}{\kilo\meter} LEO orbit, but by on-board cryptographic operations. 
Four ECC-P256 signatures across the dual-SE architecture (two identity, two attestation) account for roughly \SI{100}{}-\SI{400}{\milli\second} depending on I\textsuperscript{2}C scheduling. 
The total per-exchange time is therefore approximately \SI{210}{}-\SI{620}{\milli\second}.

The total time to certificate issuance, however, is governed by
orbital mechanics rather than cryptographic cost. 
Accumulating the required $\adversarialGS + 2 \cdot \adversarialChannels + 1$ endorsements from
geographically distributed \glspl{gs} typically requires 4-7 orbital passes in LEO, corresponding to roughly \SI{6}{}-\SI{11}{\hour} of wall-clock time for a sparse deployment, well within a $\adversarialChannels$ window of 12 hours. 
For GEO satellites, where \glspl{gs} are permanently visible, the process completes in a single session. Once $\mathit{Cert}_S$ is issued, subsequent
attestation requires only a fresh dual-signed EAT token (${\sim}$\SI{100}{}-\SI{200}{\milli\second}), making \gls{seap} a one-time setup cost.
Bandwidth overhead is negligible: each exchange transfers approximately
\SI{1.9}{\kilo\byte} under the current ECC-based instantiation. 
A migration to post-quantum signatures (e.g., Falcon-512 or ML-DSA) would increase the \textit{hello-ack} payload to approximately \SI{6}{}-\SI{12}{\kilo\byte} per exchange, which remains well within LEO link capacity. 
A detailed breakdown of latency, bandwidth, and post-quantum impact is provided
in Appendix~\ref{app:eval}.

\subsection{Radiation Resilience and Fault Tolerance}
\label{subsec:radiation-resilience}

The orbital radiation environment (\Cref{app:radiation}) introduces a fault model in which \glspl{seu} can silently corrupt cryptographic keys or execution state, and \glspl{sel} can permanently disable hardware components. 
Space Fabric mitigates both through a layered defense spanning physical, hardware, and software measures, calibrated to the power and mass budget of the satellite platform. 

\paragraph{Physical mitigation.} The payload uses standard aluminum enclosure shielding for the primary bus, augmented by localized spot-shielding placed directly over critical semiconductor junctions, reducing the particle flux reaching sensitive die areas and extending the \gls{tid} lifetime of \gls{cots} components.

\paragraph{Hardware redundancy.} To survive \glspl{sel} and isolate hardware failures, we rely on dual USB Armory Mk~II units, dual \glspl{se}, and two independent carrier boards.

\paragraph{Software integrity and execution redundancy.} At the software layer, several complementary mechanisms protect against \gls{seu}-induced corruption, again scaled to the platform's computational budget using storage redundancy and regular checkpointing with on-Earth infrastructure. We are exploring future extensions through threshold signing and computation replication to further strengthen the system. 

\section{Discussion}
\label{subsec:discussion}

\paragraph{Multi-Satellite Topologies and GEO Applicability.} The current design focuses on a single LEO satellite endorsed by Earth-based \glspl{gs}, but \gls{seap} generalizes to constellations where attested satellites endorse newly launched peers. 
A satellite holding a valid $Cert_S$ can play the role of $GS_i$: it sends a signed \texttt{hello}, appraises the \gls{eat} token, and issues an endorsement to the original \gls{gs} quorum.
The newcomer accumulates endorsements from a mix of \glspl{gs} and peer satellites until it reaches the threshold. 
The architecture also applies to \gls{geo} deployments. 
A \gls{geo} satellite has permanent visibility over a fixed set of \glspl{gs}, so $\adversarialChannels$ can be tightened to minutes and endorsements collected in a single session rather than across multiple orbital periods. 
The protocol requires no structural changes for any of these topologies, though the verifier must validate the endorsing satellite's own $Cert_S$ against the current \gls{gs} committee, creating a transitive trust chain whose security depends on the resistance to posterior corruptions property. 
The \texttt{hello}, \texttt{hello-ack}, and \texttt{key-verify} messages are agnostic to whether the endorser is a ground station or peer satellite.

\paragraph{Availability, Radiation, and Fault Tolerance.}
The current \gls{seap} instantiation uses a single USB Armory Mk~II as the sole \gls{tee}-capable platform aboard the satellite, representing a single point of failure for both attestation and identity operations. 
This is also a problem for ionizing radiation.
A \gls{seu} flipping a single bit in an attestation key, a nonce, or a signature share is functionally equivalent to a fault-injection attack on a terrestrial \gls{tee}~\cite{faultinjection_tee_mobile,radeffects_advanced,mehlitz2005radhard_sw}.
Space Fabric plans to address this through a layered mitigation strategy described in \Cref{subsec:radiation-resilience}, combining physical shielding, hardware redundancy, and software-level defenses including threshold signing and computation replication. 
This design would also strengthen the security argument: an adversary would need to compromise multiple physically separated modules simultaneously to forge attestation evidence.

\paragraph{Post-Quantum Migration.}
The current cryptographic instantiation uses \gls{ecc} for attestation signing, identity binding, and \gls{gs} endorsements,  which is vulnerable to a quantum computer. 
Migrating \gls{seap} to post-quantum primitives is, in most respects, straightforward. 
The used \gls{tpm} already supports \gls{pqc} for its firmware upgrades, serving as an important stepping stone. 
The \gls{eat} token structure is algorithm-agnostic, the endorsement and $Cert_S$ formats can accommodate any signature scheme, and the Veraison verifier supports pluggable cryptographic backends. 
The primary constraint is the \gls{hsm}, which supports now mainly \gls{ecc} and cannot be reconfigured. 
A \gls{pqc}-capable \gls{se} would be required for a fully post-quantum instantiation. 
Alternatively, the \gls{se} could retain its role as a hardware binding anchor while a software layer running inside the TrustZone Secure World handles \gls{pqc} signing, accepting the trade-off that the \gls{pqc} private key would be stored in secured on-chip SRAM rather than in a non-exportable hardware slot. 
Given the known harvest-now-decrypt-later threat applicable to long-lived attestation evidence, planning for a hybrid classical/\gls{pqc} is relevant for deployments. 
Nevertheless, other challenges with PQC, e.g., message size, remain. 

\paragraph{Software Verifiability.}
While the hardware trust anchors in \gls{seap} are designed to be independently auditable, the software stack running in the TrustZone Secure World remains a significant challenge for verification. The GoTEE Trusted OS and the \gls{pta} constitute the \gls{tcb} that produces attestation evidence, and any vulnerability in this code directly undermines the guarantees \gls{seap} provides. 
Formal verification of the \gls{pta} would provide stronger assurance than testing alone, and is particularly important given that the \gls{pta} handles all security-critical operations. Open-sourcing the full Secure World software stack is a necessary prerequisite for independent audits, and we intend to release the GoTEE \gls{pta} implementation as part of the \gls{seap} artifacts. 

\paragraph{Performance vs. Security Trade-offs.}
Orbital physical inaccessibility closes the threat vector that motivates several expensive terrestrial protections, notably off-chip bus encryption and DRAM memory encryption, which incur non-trivial power and compute overhead on resource-constrained satellite platforms. 
For instance, the I$^2$C bus between the SoC and the SE050 is unencrypted, which would be unacceptable on a terrestrial platform where bus probing is trivial but is defensible on a sealed satellite bus post-launch.
However, this reasoning must be scoped carefully. 
First, the pre-launch integration window, during which hardware is accessible to supply-chain actors, demands the full set of protections regardless of eventual deployment context. 
Second, software-based side-channel attacks do not require physical access and are not mitigated by orbital isolation. 
The appropriate posture is therefore to treat physical inaccessibility as a supplement to hardware security features, not a substitute, and to document explicitly which protections are relaxed and under what threat model assumptions.

\paragraph{Atmospheric Re-entry and Supply Chain Termination.}
A notable and often overlooked property of satellite deployments is that end-of-life re-entry into Earth's atmosphere provides a physically enforced and irreversible destruction mechanism for all on-board hardware, including the secure element, \gls{otp} fuse state, and any keying material that survived the operational lifetime. 
This stands in sharp contrast to terrestrial \gls{tee} deployments, where decommissioned hardware can be recovered, forensically analysed, and potentially used to extract residual key material or replay historical attestation evidence. 
Space re-entry effectively eliminates an entire class of supply-chain and end-of-life concerns that must be carefully managed in terrestrial deployments, as the satellite's final disposal is itself a security primitive. 
For missions using controlled de-orbit manoeuvres, the timing of this destruction event is known and predictable, enabling the operator to revoke relevant certificates. 

\subsection{Applications}
The decentralized satellite networks motivating this work~\cite{seoyual2024hotnets} require a trust foundation before multi-operator orbital services become viable. \gls{seap} and confidential computing in space provide exactly this foundation, enabling a range of services that currently require unconditional trust in the satellite operator or are avoided altogether due to the inability to verify on-board execution integrity:

\textbf{Secure multi-tenant edge computing}, where multiple customers deploy proprietary algorithms (e.g., inference models, compression pipelines) on a shared satellite, each assured that their workload runs in an isolated, attested \gls{tee} and that neither the operator nor co-tenants can observe their code or data. In a decentralized constellation, this enables operators to offer compute-as-a-service to third parties who need not trust any single operator.

\textbf{Sovereign data processing}, enabling governments or regulated industries to enforce data-residency and need-to-know policies by processing sensitive Earth-observation or signals-intelligence data on-board within an attested enclave, downlinking only derived products rather than raw imagery. Orbital attestation ensures compliance is cryptographically verifiable rather than contractually asserted.

\textbf{Secure key management and relay for inter-satellite links}, using attested \glspl{tee} as orbital key distribution nodes whose integrity ground-based verifiers can confirm, supporting encrypted mesh networking across constellations. For decentralized networks where satellites belong to different operators, \gls{seap}-attested key distribution removes the need for a shared terrestrial \gls{kms}.

\textbf{Verifiable scientific data provenance}, allowing research agencies to certify that climate, weather, or astronomical observations were processed by a known, unmodified software stack, strengthening the evidentiary value of space-derived datasets.

While the current implementation targets a specific Trusted Applet, these use cases benefit from a more general-purpose execution model. WaTZ demonstrates that WebAssembly runtimes can be hosted inside ARM TrustZone with full remote attestation support, enabling operators to deploy portable, sandboxed WASM modules as attested workloads~\cite{menetrey2022watz}. Integrating such a runtime into the GoTEE Secure World would allow third-party payloads to be uploaded, measured, and attested through the same \gls{seap} pipeline without requiring bespoke native \glspl{ta} for each application — significantly lowering the barrier to multi-tenant and updateable confidential computing in orbit.

\section{Conclusion}
As computation moves to orbit and decentralized satellite networks take shape, the ability to establish trust in remote, multi-operator orbital infrastructure becomes a prerequisite rather than an aspiration.
This paper presented Space Fabric, a satellite-native trusted computing architecture that addresses the two fundamental gaps in terrestrial \gls{tee} deployments: the physical access gap, where hardware accessible to adversaries will eventually be compromised, and the root-of-trust gap, where attestation chains depend on manufacturer-provisioned secrets. 
Space Fabric closes the first gap by relocating the trusted computing stack to orbital infrastructure, exploiting post-launch physical inaccessibility as a first-class security primitive. 
It closes the second through fully on-orbit key genesis: all cryptographic signing keys are generated within co-located secure elements after launch, with no persistent signing secrets existing on Earth at any point, a property that shifts the trust assumption from vendor secret management to operator device registration, a more auditable and operationally controllable requirement in comparison to current \gls{tee} platforms.
A dual-SE architecture, pairing the closed-source NXP SE050 with the fully auditable Tropic Square TROPIC01, eliminates single-vendor trust dependence by requiring both elements to co-sign attestation evidence.

The \gls{seap} binds these components together, cryptographically certifying not only what program executes inside the \gls{tee} but also where, through a Byzantine-tolerant ground station endorsement quorum. 
We implemented Space Fabric on a USB Armory Mk II with ARM TrustZone, demonstrated end-to-end attestation via Veraison, and established both security satisfaction arguments and impossibility bounds under a strong adaptive adversary.

Open directions include threshold cryptography for hardware redundancy, post-quantum migration, formal verification of the Secure World stack, and integration of general-purpose runtimes such as WaTZ to enable multi-tenant confidential computing in orbit.
Space Fabric demonstrates that the orbital environment's unique properties, physical inaccessibility, deterministic trajectories, and atmospheric destruction at end-of-life are exploitable security primitives that open a new design axis for trusted computing.

\bibliographystyle{ACM-Reference-Format}
\bibliography{lit.bib}

\appendix
% \begin{acks}
% This paper was edited for grammar using ChatGPT.
% \end{acks}
% \section*{Ethical Considerations}
% \textbf{This work does not raise any ethical issues. It does not involve human subjects, personally identifiable information, sensitive data, or potential misuse concerns. No experiments were conducted on individuals, communities, or systems in ways that could lead to harm.}
% % \section*{Open Science}
% % \section*{AI Usage}
% This paper was edited for grammar using ChatGPT and Grammarly.

\section{\gls{seap} Pseudocode}
This appendix contains the Algorithms \ref{alg:satellite-protocol-sat} and \ref{alg:satellite-protocol-gs}.

\label{appendix:pseudocode}
\begin{algorithm}[htbp]
    \caption{The algorithm run by a satellite $\satellite$ for the \gls{seap}.}
    \label{alg:satellite-protocol-sat}

    \KwIn{Keys $\mathbb{K} = \{ \pubkey_\gs | \gs \in \gsSet \}$, number of adversarial \glspl{gs} $\adversarialGS$, number of corrupted channels $\adversarialChannels$, channel corruption window $\channelCorruptionWindow$}

    \fn{$\texttt{sat\_satJoin}_{\secparam}(\mathbb{K}, \adversarialGS, \adversarialChannels, \channelCorruptionWindow)$}{
        $\langle \pubkey_\satellite, \privkey_\satellite \rangle \gets \keygen(1^\secparam)$

        $\teeProof^\pubkey \leftarrow \texttt{prove}(\pubkey_\satellite)$
        \tcp{TEE quote proving that $\pubkey_\satellite$ was created in the TEE}

        $\certificate_\satellite \leftarrow \bot$,
        $L \leftarrow \{ \}$

        \While{$\certificate_\satellite = \bot$}{
            $\langle \msf{msg}, \msf{data}, \gs \rangle \leftarrow \texttt{network}()$
            \tcp{Wait until message is received}

            \If{$\msf{msg} = \text{hello}$}{
                $\langle \nonce, \signature_{\gs, \text{hello}} \rangle \leftarrow \msf{data}$

                \If{$\verify(\pubkey_\gs, \langle \nonce, \text{hello} \rangle, \signature_{\gs, \text{hello}}) = 1$}{
                    $\signature_{\satellite, \nonce} \leftarrow \sign(\nonce, \privkey_\satellite)$

                    $M \leftarrow \langle \nonce, \pubkey_\satellite, \signature_{\satellite, \nonce}, \teeProof^\pubkey \rangle$

                    $\langle \text{hello-ack}, M \rangle \rightarrow \texttt{network}(\gs)$
                }
            }

            \ElseIf{$\msf{msg} = \text{key-verify}$}{
                $\langle \timestamp, \signature \rangle \leftarrow \msf{data}$

                \If{$\verify(\pubkey_\gs, \langle \pubkey_\satellite, \timestamp \rangle, \signature) = 1$}{
                    $L \leftarrow L \cup \langle \timestamp, \gs, \signature \rangle$

                    \If{exists $L' \subseteq L$ such that $(\forall s, s' \in L': | s.\timestamp - s'.\timestamp | < \channelCorruptionWindow)$ and $(| \{ s.\gs | s \in L' \} | = \adversarialGS + 2 \cdot \adversarialChannels + 1)$}{
                        \label{alg-line:time-window-check}

                        $\certificate \leftarrow \langle \pubkey_\satellite, \{ (s.\signature, s.\gs) | s \in L' \} \rangle$

                        $\certificate_\satellite \leftarrow \langle \certificate, \sign(\certificate, \privkey_\satellite) \rangle$
                    }
                }
            }
        }

        \ForEach{$\gs \in \gsSet$}{
            $\langle \text{cert}, \certificate_\satellite \rangle \rightarrow \texttt{network}(\gs)$
        }

        \Return $\certificate_\satellite$
    }
\end{algorithm}

\begin{algorithm}[htbp]
    \caption{
        The algorithm is run by a \gls{gs} $\gs$ for the \emph{satellite identification protocol}.
    }
    \label{alg:satellite-protocol-gs}

    \KwIn{Keys $\mathbb{K} = \{ \pubkey_\gs | \gs \in \gsSet \}$, channel corruption window $\channelCorruptionWindow$}

    \fn{$\texttt{GS\_satJoin}_{\secparam, \privkey_\gs}(\mathbb{K}, \channelCorruptionWindow)$}{
        $\certificate_{\satellite} \leftarrow \bot$

        \While{$\certificate_{\satellite} = \bot$}{
            $\langle \msf{satNear}, \satellite \rangle \leftarrow \texttt{env}()$
            \tcp{Wait until $\satellite$ is in range for $T$ seconds}

            $\nonce \xleftarrow{\$} \{ 0, 1 \}^\secparam$

            $\signature_\gs \leftarrow \sign(\langle \nonce, \text{hello} \rangle, \privkey_\gs)$

            $\langle \text{hello}, \langle \nonce, \signature_\gs \rangle \rangle \rightarrow \texttt{network}(\satellite)$

            $\timestamp \leftarrow \texttt{clock}()$

            \While{True}{
                $\langle \msf{msg}, \msf{data}, \satellite \rangle \leftarrow \texttt{network}()$
                \tcp{Wait until message is received}

                \If{$\texttt{clock}()-\timestamp > \channelCorruptionWindow$}{
                    \tcp{Session lasts at most $\channelCorruptionWindow$ seconds}
                    \break{}
                }
                \If{$\msf{msg} = \text{hello-ack}$}{
                    $\langle \nonce, \pubkey_\satellite, \signature_\satellite, \teeProof^\pubkey \rangle \leftarrow \msf{data}$

                    \If{$(\verify(\pubkey_\satellite, \nonce, \signature_\satellite) = 1)$ and $(\texttt{teeValidate}(\teeProof^\pubkey))$}{
                        $\timestamp' \leftarrow \texttt{clock}()$

                        $\signature \leftarrow \sign(\langle \pubkey_\satellite, \timestamp' \rangle, \privkey_\gs)$

                        $\langle \text{key-verify}, \langle \timestamp', \signature \rangle \rangle \rightarrow \texttt{network}(\satellite)$
                    }
                }
                \ElseIf{$\msf{msg} = \text{cert}$}{
                    $\langle \certificate, \signature \rangle \leftarrow \msf{data}$

                    $\langle \pubkey_\satellite, \mathbb{S} \rangle \leftarrow \certificate$
                    
                    \If{$(\forall (s.\signature, s.\gs) \in \mathbb{S}: \verify(\pubkey_{s.\gs}, \pubkey_\satellite, s.\signature) = 1)$ and $(\verify(\pubkey_\satellite, \certificate, \signature) = 1)$}{
                        $\certificate_\satellite \leftarrow \langle \certificate, \signature \rangle$

                        \break{}
                    }
                }
            }
        }

        \Return $\certificate_\satellite$
    }
\end{algorithm}

\section{Satellite Computing Extended}
This appendix provides additional details on space deployments and the challenges they pose. 
\subsection{Satellite Deployment Options}

Space Fabric's orbital infrastructure can be deployed through two primary approaches, each offering distinct trade-offs between cost, capability, and deployment timeline.

\paragraph{CubeSat Platforms} CubeSats are standardized small satellites based on 10×10×10 cm units, with typical 3U or 6U configurations providing sufficient volume for computing payloads, communication transceivers, and power systems at development costs of \$500K-\$2M per satellite~\cite{poghosyan2017cubesat,sweeting2018modern}. However, CubeSats face power constraints (30-50W average), shorter operational lifetimes (2-5 years in LEO), and limited computational capability compared to larger satellites~\cite{shiroma2021cubesats}.

\paragraph{Rideshare Launch Services} Rideshare missions allow multiple satellites to share launch costs, with services like SpaceX's Transporter offering deployment at approximately \$275K per 200 kg, dramatically cheaper than dedicated launches~\cite{spacex_rideshare}. The primary trade-off is limited control over orbital parameters and launch timing, as satellites must accept the primary mission's target orbit and schedule~\cite{crisp2020small}.

\paragraph{Space Fabric Deployment Strategy} A practical deployment would use an initial 3U-8U CubeSat demonstration mission via rideshare (\$300K-500K total), followed by a production constellation of 4-8 CubeSats for global coverage, with total launch costs of \$400K-600K across multiple rideshare missions. Ground station services from commercial providers (AWS Ground Station, KSAT) eliminate the need for dedicated infrastructure at \$3-10 per minute of contact time~\cite{aws_ground_station}.

\subsubsection{Latency and Bandwidth Considerations}

The security advantages of satellite infrastructure come with performance trade-offs dominated by communication latency until solutions like Starlink become widely available. 

\paragraph{\gls{leo} Satellites} Operating at altitudes of 500-2000 km, \gls{leo} satellites provide the most favorable latency characteristics for interactive applications. 
Signal propagation at the speed of light imposes a fundamental lower bound: a signal traveling 1000 km to a satellite and 1000 km back incurs approximately 7ms of propagation delay. 
In practice, LEO satellite round-trip latency ranges from 20-40ms one-way when accounting for atmospheric propagation, ground station processing, and on-board routing delays~\cite{shiroma2021cubesats}.
This latency is comparable to wide-area network communication between distant data centers, making LEO satellites suitable for applications tolerating moderate latency—including periodic attestation, key derivation, and execution monitoring where interactions occur on second-to-minute timescales rather than milliseconds.
\gls{geo} satellites impose higher latency: the signal path length of approximately 71,572 km (up and down) translates to roughly 250ms of propagation delay when including ground and satellite processing~\cite{shiroma2021cubesats}. 

\paragraph{Bandwidth Constraints} Modern LEO constellations support throughput ranging from several Mbps to hundreds of Mbps per user terminal, while GEO satellites provide tens to hundreds of Mbps depending on frequency band and antenna configuration~\cite{kodheli2021satellite}. 

\paragraph{Design Implications} Space Fabric's protocol design explicitly accounts for satellite latency. 
Rather than requiring synchronous satellite communication for every operation, the system batches attestation requests, caches satellite-derived keys with limited lifetimes, and maintains local execution state.
This design enables ground platforms to operate with reasonable performance during normal operation while maintaining security properties that depend on satellite verification. 
For applications like batch processing, federated learning, edge analytics, and secure enclaves processing sensitive data over minutes to hours, the LEO satellite latency overhead (40-80ms per attestation cycle) represents an acceptable trade-off for physical security guarantees.

\subsection{Space Radiation Phenomena}
\label{app:radiation}

Orbital platforms operate in a hard ionizing radiation environment that may impact semiconductor reliability. 
Unlike traditional space missions that rely on highly specialized and computationally limited radiation-hardened (Rad-Hard) processors, the emerging space computing paradigm heavily leverages \gls{cots} silicon to meet the Size, Weight, and Power (SWaP) constraints of modern small satellites~\cite{radeffects_advanced,mehlitz2005radhard_sw}. 
Consequently, space computing architectures must account for two primary radiation phenomena: cumulative dose effects and transient single-event effects.

\gls{tid} refers to the long-term degradation of semiconductor materials caused by continuous exposure to trapped protons and electrons in the Earth's radiation belts.
Over time, this ionizing radiation alters transistor threshold voltages, leading to increased leakage current and eventual permanent device failure. 
For \gls{cots} components in typical \gls{leo} deployments, \gls{tid} is the principal determinant of hardware payload operational lifetime.

\glspl{see} pose a more immediate threat to continuous computation. 
They occur when a single high-energy particle — such as a heavy ion from Galactic Cosmic Rays or a trapped proton from the South Atlantic Anomaly — strikes the silicon and deposits a localized charge~\cite{radeffects_advanced}. 
\glspl{see} fall into two broad categories. 
\glspl{seu}, commonly referred to as bit flips, are non-destructive transient faults in which the state of a memory cell (SRAM or DRAM) or logic register is inverted. 
In the context of a \gls{tee}, an unmitigated \gls{seu} is functionally equivalent to an arbitrary fault-injection attack~\cite{faultinjection_tee_mobile}: a flipped bit in a cryptographic key, an attestation nonce, or a program counter can silently corrupt execution bounds or invalidate digital signatures. 
\glspl{sel} are more severe: a particle strike triggers a parasitic thyristor structure within the CMOS silicon, creating a low-impedance path to ground. 
\glspl{sel} disrupt device operation and cause high current draw, potentially leading to thermal destruction if the affected component is not rapidly power-cycled.

Space Fabric's mitigation strategy for both phenomena is described in \Cref{subsec:radiation-resilience}.
The key observation for the security architecture is that \glspl{seu} must be treated as an adversarial fault model in which any single bit in the \gls{tee}'s memory or register file may be flipped at any time, and the system must detect or tolerate such faults without silently producing corrupted cryptographic output.

\section{Protocol Evaluation}\label{app:eval}

This appendix provides a detailed breakdown of \gls{seap}'s performance characteristics under realistic deployment parameters, covering per-exchange latency, total time to certificate issuance, bandwidth overhead, and the impact of a post-quantum cryptographic migration.

\subsection{Per-Ground-Station Exchange Latency}

Each \gls{seap} exchange with a single ground station $\mathcal{GS}_i$
consists of three one-way messages: \texttt{hello} ($\mathcal{GS} \to \mathcal{S}$), \texttt{hello-ack} ($\mathcal{S} \to \mathcal{GS}$), and \texttt{key-verify} ($\mathcal{GS} \to \mathcal{S}$), totaling 1.5 round-trip. \Cref{tab:latency} breaks down the expected latency contributions.

\begin{table}[t]
\centering
\caption{Per-exchange latency breakdown (500\,km LEO).}
\label{tab:latency}
\small
\begin{tabular}{@{}lr@{}}
\toprule
\textbf{Component} & \textbf{Duration} \\
\midrule
Propagation ($3\times$ one-way)\textsuperscript{a} & 60--120\,ms \\
On-board crypto (dual-SE)\textsuperscript{b} & 200--400\,ms \\
GS-side verification\textsuperscript{c} & 50--100\,ms \\
\midrule
\textbf{Total (sequential)} & \textbf{310--620\,ms} \\
\textbf{Total (parallel SEs)} & \textbf{210--420\,ms} \\
\bottomrule
\multicolumn{2}{l}{\textsuperscript{a}20--40\,ms one-way incl.\ atmospheric and processing delays~\cite{kodheli2021satellite}.} \\ 
\multicolumn{2}{l}{\textsuperscript{b}4$\times$ ECC-P256 sign over I\textsuperscript{2}C; parallel assumes independent buses.} \\ 
\multicolumn{2}{l}{\textsuperscript{c}2$\times$ signature verify + Veraison appraisal + timing check.} \\
\end{tabular}
\end{table}
The four on-board signing operations: $\sigma_{S,r_i}$ and
$\sigma'_{S,r_i}$ (identity keys, slots~0) and $\pi^{\mathit{TEE}}_{\mathit{NXP}}$
and $\pi^{\mathit{TEE}}_{\mathit{TROP}}$ (attestation keys, slots~1), are the primary bottleneck. 
With typical ECC-P256 signing latency of \SI{50}{}-\SI{100}{\milli\second} per operation on embedded secure elements, this yields approximately \SI{100}{}-\SI{200}{\milli\second} under parallel scheduling.

\subsection{Total Time to Certificate Issuance}

\gls{seap} requires $\adversarialGS + 2 \cdot \adversarialChannels + 1$ valid endorsements within $\adversarialChannels$. 
The total time is dominated by orbital mechanics.~\Cref{tab:cert-time} parametrizes this under conservative and moderate adversarial assumptions.

\begin{table}[h]
\centering
\caption{Estimated time to $\mathit{Cert}_S$ issuance (500\,km LEO, 95\,min orbital period).}
\label{tab:cert-time}
\begin{tabular}{lcc}
\toprule
\textbf{Parameter} & \textbf{Conservative} & \textbf{Moderate} \\
\midrule
$\adversarialGS$ (adversarial GSs) & 2 & 3 \\
$\adversarialChannels$ (corrupted channels) & 2 & 3 \\
Required endorsements & 7 & 10 \\
GS contacts per orbit & 1--2 & 2--3 \\
Orbits to completion & 4--7 & 4--5 \\
\textbf{Wall-clock time} & \textbf{6--11\,h} & \textbf{6--8\,h} \\
$\Delta_\mathit{corr}$ (configured) & 12\,h & 12\,h \\
\bottomrule
\end{tabular}
\end{table}

For a sparse \gls{gs} deployment (8--12 globally distributed stations), the satellite typically contacts 1--3 stations per orbit depending on orbital inclination and station placement. Denser deployments or higher-inclination orbits reduce the time proportionally. 
For GEO satellites, where a fixed set of \glspl{gs} is permanently visible, the entire endorsement collection can be completed in a single session, potentially within minutes, since no orbital passes need to be awaited.

\subsection{Bandwidth Overhead}

Table~\ref{tab:bandwidth-ecc} summarizes the per-exchange bandwidth under the current ECC-P256 instantiation.

\begin{table}[h]
\centering
\caption{Bandwidth per SEAP exchange (ECC-P256 instantiation).}
\label{tab:bandwidth-ecc}
\begin{tabular}{lrl}
\toprule
\textbf{Message} & \textbf{Size} & \textbf{Direction} \\
\midrule
\texttt{hello} & $\sim$\SI{200}{\byte} & $\mathcal{GS} \to \mathcal{S}$ \\
\texttt{hello-ack} & $\sim$\SI{1.5}{\kilo\byte} & $\mathcal{S} \to \mathcal{GS}$ \\
\texttt{key-verify} & $\sim$\SI{150}{\byte} & $\mathcal{GS} \to \mathcal{S}$ \\
\midrule
\textbf{Total per exchange} & \textbf{$\sim$\SI{1.9}{\kilo\byte}} & \\
% \textbf{Total for certificate} (10 exchanges) & \textbf{$\sim$19\,KB} & \\
\bottomrule
\end{tabular}
\end{table}

This is negligible relative to the bandwidth available on modern LEO links (in the Mbps range) and well within the capacity of limited S-band channels. 
The certificate $\mathit{Cert}_S$ itself, a bundle of endorsement signatures plus the \gls{tee}'s own signature, adds approximately \SI{2}{}-\SI{3}{\kilo\byte} as a one-time broadcast.

\subsection{Post-Quantum Bandwidth Impact}
\label{app:pqc-eval}

A migration to post-quantum cryptography substantially increases the sizes of signatures and keys while leaving the protocol structure unchanged. 
Table~\ref{tab:pqc-sizes} compares the relevant primitive sizes across the current ECC instantiation and two PQC candidates: Falcon-512 for signatures \cite{NIST-FIPS206,Falcon-spec} and ML-KEM-768 for key encapsulation \cite{NIST-FIPS203}.

\begin{table}[h]
\centering
\caption{Cryptographic primitive sizes: ECC-P256 vs.\ post-quantum candidates.}
\label{tab:pqc-sizes}
\begin{tabular}{lccc}
\toprule
\textbf{Primitive} & \textbf{ECC-P256} & \textbf{Falcon-512} & \textbf{ML-KEM-768} \\
\midrule
Public key & \SI{64}{\byte} & \SI{897}{\byte} & \SI{1184}{\byte} \\
Signature & \SI{64}{\byte} & \SI{666}{\byte} & \SI{1088}{\byte} \\
Secret key & \SI{32}{\byte} & \SI{1281}{\byte} & \SI{2400}{\byte} \\
\bottomrule
\end{tabular}
\end{table}

Table~\ref{tab:pqc-bandwidth} projects the impact on the message sizes \gls{seap}
 under a hybrid ECC+Falcon signature scheme (retaining ECC for backward compatibility) and a complete migration to Falcon-only. 
We also consider the case where a secure channel between the verifier and satellite is established post-certificate using ML-KEM-768 for key encapsulation.

\begin{table}[h]
\centering
\caption{SEAP bandwidth per exchange under post-quantum migration.}
\label{tab:pqc-bandwidth}
\begin{tabular}{lccc}
\toprule
\textbf{Message} & \textbf{ECC-only} & \textbf{Hybrid} & \textbf{Falcon-only} \\
&  & \textbf{(ECC+Falcon)} &  \\
\midrule
\texttt{hello} & $\sim$ \SI{200}{\byte} & $\sim$ \SI{930}{\byte} & $\sim$\SI{730}{\byte} \\
\texttt{hello-ack} & $\sim$ \SI{1.5}{\kilo\byte} & $\sim$ \SI{6.2}{\kilo\byte} & $\sim$ \SI{5}{\kilo\byte} \\
\texttt{key-verify} & $\sim$ \SI{150}{\byte} & $\sim$\SI{880}{\byte} & $\sim$ \SI{730}{\byte} \\
\midrule
\textbf{Total} & \textbf{$\sim$\SI{1.9}{\kilo\byte}} & \textbf{$\sim$\SI{8}{\kilo\byte}} & \textbf{$\sim$\SI{6.5}{\kilo\byte}} \\
% \textbf{Total for certificate} (10 ex.) & \textbf{$\sim$19\,KB} & \textbf{$\sim$80\,KB} & \textbf{$\sim$65\,KB} \\
\bottomrule
\end{tabular}
\end{table}

The \texttt{hello-ack} message is the most affected, as it carries two public keys, two attestation proofs (each containing a Falcon signature), and two nonce signatures.
Under the hybrid scheme, the \texttt{hello-ack} grows from $\sim$\SI{1.5}{\kilo\byte} to $\sim$\SI{6.2}{\kilo\byte}, a $4\times$ increase. 
This remains well within \gls{leo} link capacity but becomes relevant for bandwidth-constrained S-band contacts, where a full certificate exchange (10 \glspl{gs}, hybrid signatures) would consume approximately \SI{80}{\kilo\byte} compared to \SI{19}{\kilo\byte} under ECC.

\paragraph{Key exchange for post-certificate channels.}
After $\mathit{Cert}_S$ is issued, a verifier establishing a secure channel with the satellite may use ML-KEM-768 for key encapsulation. 
This adds \SI{1184}{\byte} (public key) + \SI{1088}{\byte} (ciphertext) $\approx$ \SI{2.3}{\kilo\byte} to the channel establishment handshake, a one-time cost per session that is negligible relative to the \gls{seap} overhead itself. 
The shared secret derived from ML-KEM can then protect subsequent communication using a symmetric cipher, with no further PQC overhead on the data path.

\paragraph{Computational impact.}
Falcon-512 signing is computationally more expensive than ECC-P256, with typical software implementations requiring 2--5$\times$ the cycle count~\cite{Howe2023-FalconARM,Becker2022-FalconARMv8}. 
On embedded secure elements, where ECC signing takes \SI{50}{}-\SI{100}{\milli\second}, a software-based Falcon implementation inside the TrustZone Secure World could require \SI{200}{}-\SI{500}{\milli\second} per signature. 
Since Falcon signing must currently be performed in software (neither the SE050 nor TROPIC01 supports PQC natively), the PQC private key would reside in secured on-chip SRAM rather than a non-exportable hardware slot. Under the hybrid scheme, each SE still signs with its hardware-bound ECC key (ensuring hardware binding), while a software layer adds the Falcon co-signature (ensuring quantum resistance). 
The per-exchange on-board crypto time would increase from $\sim$\SI{100}{}-\SI{400}{\milli\second} to approximately \SI{400}{}-\SI{1200}{\milli\second}, which remains acceptable for a one-time setup protocol.

ML-KEM-768 encapsulation and decapsulation are computationally lightweight $<$\SI{1}{\milli\second} on ARM Cortex-A7 class processors), so the post-certificate key exchange adds negligible latency~\cite{Becker2022-PQTLS,Paquin2020-PQTLS}.

\subsection{Operational Summary}

\begin{table}[h]
\centering
\caption{Summary of SEAP performance characteristics.}
\label{tab:eval-summary}
\begin{tabular}{lcc}
\toprule
\textbf{Metric} & \textbf{ECC-P256} & \textbf{Hybrid} \\
&  & \textbf{(ECC+Falcon)} \\
\midrule
Per-exchange latency & \SI{210}{}-\SI{410}{\milli\second} & \SI{400}{}-\SI{1200}{\milli\second} \\
Time to $\mathit{Cert}_S$ (LEO, 10 GSs) & 6--11\,h & 6--11\,h \\
Time to $\mathit{Cert}_S$ (GEO) & minutes & minutes \\
Bandwidth per exchange &  \SI{1.9}{\kilo\byte} &  \SI{8}{\kilo\byte} \\
% Total bandwidth for certificate & 19\,KB & 80\,KB \\
Post-cert attestation latency & \SI{100}{}-\SI{200}{\milli\second} & \SI{300}{}-\SI{700}{\milli\second} \\
% Post-cert channel setup & --- & +2.3\,KB one-time \\
On-orbit key genesis & $<$1\,s (once) & $<$1\,s (once) \\
\bottomrule
\end{tabular}
\end{table}

The \gls{seap} imposes modest per-exchange costs, dominated by embedded cryptographic operations rather than network latency.
The total time to certificate issuance is determined by orbital mechanics and ground-station density, not by protocol overhead. 
A post-quantum migration increases bandwidth by ${\sim}4\times$ and per-exchange latency by ${\sim}2$-$3\times$ under the hybrid scheme, both of which remain within operational bounds for LEO and GEO deployments. 
The key constraint for PQC adoption is not protocol performance but \gls{se} support: until PQC-capable \glspl{hsm} are available, the hybrid approach of hardware-bound ECC for platform binding with software Falcon for quantum resistance, offers a practical migration path.

\printglossary

\begin{acks}
We would like to thank the Common Prefix team for their support with the paper. 
\end{acks}

\end{document}